\RequirePackage{fixltx2e}
\documentclass{style}
\usepackage{subfigure}
\usepackage{bm}
\usepackage{amsmath, amssymb}
\usepackage{mathrsfs}
\usepackage{mathptmx}
\usepackage{upgreek}
\usepackage[utf8]{inputenc}
\usepackage{makecell}
\usepackage{fixltx2e}
\begin{document}

\title{A vertex-weighted-Least-Squares gradient reconstruction}

\author{Fan Zhang \\State Key Laboratory of Structural Analysis for Industrial Equipment, \\ Dalian University of Technology, People's Republic of China \\ \textit{a04051127@mail.dlut.edu.cn}}

\begin{abstract}
\noindent  Gradient reconstruction is a key process for the spatial accuracy and robustness of finite volume method,
especially in industrial aerodynamic applications in which grid quality affects reconstruction methods significantly.
A novel gradient reconstruction method for cell-centered finite volume scheme is introduced.
This method is composed of two successive steps.
First, a vertex-based weighted-least-squares procedure is implemented to calculate vertex gradients, and then the cell-centered gradients are calculated by an arithmetic averaging procedure.
By using these two procedures, extended stencils are implemented in the calculations, and the accuracy of gradient reconstruction is improved by the weighting procedure.
In the given test cases, the proposed method is showing improvement on both the accuracy and convergence. Furthermore, the method could be extended to the calculation of viscous fluxes.
\end{abstract}

\maketitle
\textbf{Key words~~~}  finite volume method, gradient reconstruction, cell-centered, vertex-based weighted-least-squares
 
\section{Introduction}

Unstructured finite volume method (FVM) has been extensively used for computational fluid dynamics (CFD) in virtue of its capability
to automatically discretize complex domains. An extensive overview of unstructured FVM discretization and solvers is given by
Mavriplis \cite{Mavriplis2007}, and thorough studies on the gradient reconstruction methods were provided by Diskin, \textit{et al}. \cite{diskin2008accuracy,Diskin2010,Diskin2011}.

For cell-centered (CC) FVM, one class of gradient reconstruction methods are (vertex-based) node-averaging-Green-Gauss (NA-GG)
schemes \cite{HOLMES1989,FRINK1991,RAUSCH1992,Frink1994,Kim2003,Katz2012}. NA-GG schemes usually perform
pseudo-laplacian (PL) procedure \cite{HOLMES1989,RAUSCH1992,Frink1994,Kim2003} to calculate vertex values at first,
and then cell gradients are calculated by GG procedure \cite{BARTH1989},
which performs the numerical integration over cell interfaces. Another class of gradient reconstruction methods are cell-based weighted-least-squares (WLSQ) methods.
In order to improve the accuracy of WLSQ method, full-augmentation (FA) stencil or smart-augmentation (SA) stencil are usually implemented \cite{diskin2008accuracy,Diskin2011}.
Newly developed WLSQ(G) scheme \cite{Shima2010,Shima2013} introduces the effects of interface area and grid distortion for improving the accuracy on unstructured grids.

In this study, a vertex-weighted-least-squares (VWLSQ) gradient reconstruction method is introduced. The method could be taken as an improvement of the PL scheme,
of which the calculations are mainly preformed at grid vertexes.
The difference is that the proposed method has utilised the vertex gradients which are discarded by PL scheme.  The presented
paper is organized as follows. In the next section, the basic of finite volume
formulation of the governing equations is given. In section \ref{sec:ProposedMethod}, the simple presented method is introduced.
Some of the characters of the presented method are introduced in section \ref{sec:Comparisons}, and the numerical comparisons are given in section \ref{sec:tests}.
Finally, the paper ends with the conclusions.

\section{Governing equations and discretization} \label{sec:GoverningEquations}

The discretization for the compressible Euler equations is introduced as follows. The integral form of the equations is
\begin{equation} \label{eq:GoverEq}
\int\limits_\Omega \frac{\partial \mathbf{Q}}{\partial t}\mathrm{d}\Omega+\int\limits_{\partial \Omega}\mathbf{F}_c(\mathbf{Q})\cdot\mathbf{n}\mathrm{d}S=0.
\end{equation}
\noindent where the $\Omega$ and $\partial\Omega$ are the control volume and its boundary surface respectively, and the $\mathbf{Q}$ is the
conservative variables and the $\mathbf{F}_c$ is the convective flux.

The governing equations are discretized using a cell-centered finite volume formulation and applied to polygon computational cell $i$,
which is sharing an interface $k$ with a neighbouring cell $j$.
Therefore, the spatial discretization at a fixed cell $i$ for the Euler equations in Eq.\ref{eq:GoverEq} can be expressed as
\begin{equation} \label{eq:SpatialDiscrete}
\frac{\partial}{\partial t}(\mathbf{Q}\Omega)_i=-(\sum_{k=1}^{N_f}\mathbf{F}_{c, k}\cdot\mathbf{n}_k S_k)_i,
\end{equation}

\noindent where $S_k$ is the interface area, $\mathbf{n}_k$ is the unit norm vector outward from the interface, and $N_f$ is the interface number of cell $i$.
The convective flux vector $\mathbf{F}_{c, k}$ could be computed according to the unified form of flux-difference-splitting (FDS) \cite{Roe1981} scheme:
\begin{equation}
\mathbf{F}_{c, k}=\frac{1}{2}[\mathbf{F}_{c}(\mathbf{Q}_{k}^+)+\mathbf{F}_{c}(\mathbf{Q}_{k}^-)-|\mathbf{\hat{A}}|(\mathbf{Q}_{k}^--\mathbf{Q}_{k}^+)].
\end{equation}

\noindent The $\mathbf{Q}_{k}^+$ and $\mathbf{Q}_{k}^-$ are the left and right variables at interface $k$ respectively.
In this study, linear reconstruction methods are employed. The cell interface variables are extrapolated from the cell-centre variables by using the gradient $\nabla \mathbf{Q}$:
\begin{equation}
\begin{split}
\mathbf{Q}_{k}^+ = \mathbf{Q}_i+ \phi_i\nabla \mathbf{Q}_i\cdot\Delta\mathbf{x}_{ik},  \\
\mathbf{Q}_{k}^- = \mathbf{Q}_j+ \phi_j\nabla \mathbf{Q}_j\cdot\Delta\mathbf{x}_{jk},
\end{split}
\end{equation}

\noindent 
where $\Delta(\cdot)_{ki}=(\cdot)_k-(\cdot)_i$.
The slope limiter coefficients \cite{BARTH1989,Venkatakrishnan1995} is usually employed to suppress oscillations at captured discontinuities.
However, the implementation of limiter will blur the accuracy estimation of gradient reconstruction,
and thus $\phi$ is set as $1$ in this paper to investigate the unlimited gradients.

\section{Proposed method} \label{sec:ProposedMethod}

A weighted-least-squares interpolation is performed at each vertex in the proposed method.
In a vertex-based gradient calculation, the cell-centre variable could be expressed as an extrapolation from the variable at a vertex of cell by using the vertex gradient,
\begin{equation} \label{eq:Expansion}
\mathbf{Q}_i=\mathbf{Q}_{l}+\nabla\mathbf{Q}_{l}\cdot\Delta\mathbf{x}_{li}.
\end{equation}

\noindent For each connected cell $i$ of vertex $l$, a linear equation as Eq.\ref{eq:Expansion} can be established,
and then an over-determined system of equations can be written as 
\begin{equation} \label{eq:OverDeterminedSystems}
\left(
  \begin{array}{ccc}
    \omega_{1}    & \omega_{1}\Delta x_{li}          & \omega_{1}\Delta y_{li}         \\
    \vdots         & \vdots                        & \vdots                       \\
    \omega_{N_l} & \omega_{N_l}\Delta x_{l N_l} & \omega_{N_l}\Delta y_{l N_l}\\
  \end{array}
\right)
\left( \begin{array}{c}  q_{l} \\ \dfrac{\partial q_{l}}{\partial{x}} \\ \dfrac{\partial q_{l}}{\partial{y}}\\ \end{array} \right)
=\left( \begin{array}{c} \omega_{1}q_{1} \\ \vdots   \\ \omega_{N_l}q_{N_l}     \end{array} \right),
\end{equation}

\noindent where the $N_l$ is the connected cell number of vertex $l$. The system can also be written as
\begin{equation}
\mathbf{A}_{l}\mathbf{X}_{l}=\mathbf{B}_{l}.
\end{equation}

The unknown vertex value $\mathbf{Q}_{l}$ and gradient $\nabla\mathbf{Q}_{l}$ can be attained by solving the over-determined system by the means of least squares method.
The $\omega_{i}$ in Eq.\ref{eq:OverDeterminedSystems} is the weight of the weighted-least-squares procedure which usually is inversely proportional to the distance
$L_{li}$ between cell $i$ and vertex $l$. The weight can be written in an universal form as
\begin{equation}
\omega_{i}=\frac{1}{L_{li}^n}.
\end{equation}

\noindent Therefore, the proposed scheme is finally named as VWLSQ($n$) for convenience.
The unweighted-vertex-least-squares method could be represented as VWLSQ($0$) scheme in the unified manner.

The proposed method is less concerned with the vertex value $\mathbf{Q}_{l}$ but focuses on the first order spatial derivative $\nabla \mathbf{Q}_{l}$.
Once the vertex gradients are calculated, the gradients in cells and interfaces can both be calculated by an arithmetical averaging procedure as
\begin{equation}\label{eq:Ave}
\begin{split}
\nabla\mathbf{Q}_{i}= \left. \sum_{l=1}^{N_v}\nabla\mathbf{Q}_{l} \middle / {N_v} \right., \\
\nabla\mathbf{Q}_{k}= \left. \sum_{l=1}^{N_v}\nabla\mathbf{Q}_{l} \middle / {N_v} \right.,
\end{split}
\end{equation}
\noindent where the $N_v$ is the adjoining vertex number of cell $i$ or interface $k$.
For the second-order spatial discretization, the estimation of gradient is expected to be first-order accuracy, or constant.
Therefore, a directly arithmetical averaging procedure of second-order accuracy is sufficient for the computation of the gradients.

\section{Explanation and comparisons} \label{sec:Comparisons}

\subsection{Vertex-based gradient reconstructions}  \label{sec:Comp1}

\begin{figure}
 \centering
 \includegraphics[width=1.8in]{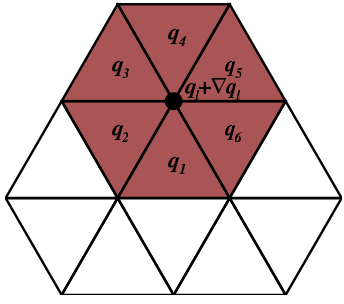}
 \caption{ \label{fig:f:stencil}
  Stencils for vertex-based reconstruction. Vertex stencils: shadowed cells. cell stencils: all cells.}
\end{figure}

The FA stencil, which includes all the common vertex neighbour cells, is a natural advantage of vertex-based reconstructions, including the
PL-CLIP (CLIPping) scheme \cite{HOLMES1989,RAUSCH1992,Frink1994,Kim2003} and the proposed VWLSQ($n$) scheme.
As shown in Fig.\ref{fig:f:stencil}, FA stencil provides flow information around the central cell.
It means the gradient reconstruction will be less affected by grid distortion and/or lack of neighbouring cells,
and thus the accuracy and stability could be retained in complex grid conditions.

Furthermore, the vertex-based reconstruction possesses the efficiency advantage.
Table \ref{tab:ComputationCosts}, in which the number $N$ is the vertex number of each grid type, is showing how many least-squares problems should be solved on each grid type.
For example, on a quadrilateral grid, cell number is approximately equal to vertex number, and
thus the computation cost of vertex-based schemes or cell-based schemes are in the same level.
However, on a triangular grid, vertex number is half of cell number, and then the vertex-based WLSQ scheme will be
more efficient. Especially, reconstructions using FA stencil require more computations for each cell.

Therefore, for inviscid flow simulations on quadrilateral and hexahedron grids,
the computation costs of cell-based or vertex-based methods would be about the same.
On triangular or tetrahedron grids, the vertex-based reconstruction methods are more efficient compared with the cell-based WLSQ scheme,
because the vertexes are much less than the cells on such grid types. For viscous flow simulations,
the facial gradient reconstruction is a necessary step for the computation of viscous fluxes \cite{Diskin2010}.
The VWLSQ($n$) scheme will be more efficient because
its facial gradients could be calculated by the simple averaging procedure based on vertex gradients.

\begin{table}
  \centering
  \caption{Requirement of LSQ solutions for gradient reconstructions}\label{tab:ComputationCosts}
  \begin{tabular}{lcccc}
   \hline
    &Triangle&Tetrahedron&Quadrangle&Hexahedron
   \\\hline
   Vertex     & $N$   & $N$    & $N$   & $N$    \\
   Cell     & $2N$  & $6N$   & $N$   & $N$    \\
   Face     & $3N$  & $12N$  & $2N$  & $3N$   \\
   \hline
  \end{tabular}
\end{table}

\subsection{Invoking of clipping pricedure} \label{sec:clip}
The PL scheme with clipping procedure, or the PL-CLIP scheme, is commonly used because its efficiency and effectiveness.
Although there was a research which suggested that the clipping procedure is unnecessary except at the boundaries \cite{Jawahar2000},
the procedure has been applied to ensure positivity of weights used to calculate vertex variables. Therefore, a brief discussion is presented here for this issue.
\begin{figure}
 \centering
 \includegraphics[height=1.5in]{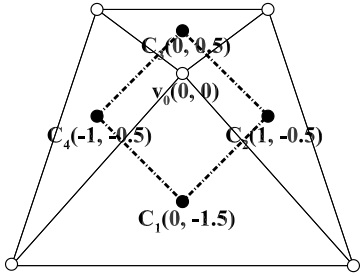}
 \caption{ \label{fig:f:no_balence}
  A critical situation of invoking clipping procedure }
\end{figure}

In Fig.\ref{fig:f:no_balence}, four triangular cells are given of which the cell centre coordinates are assigned.
The value at vertex $v_0$ is unknown and needs to be calculated by an interpolation of four cell centre values. If the PL scheme \cite{RAUSCH1992} is applied, the weight of cell centre $C_1$, $\omega_{01}$, is zero.
Therefore, this is a critical situation that may be leading to a negative weight, and thus the clipping procedure will be used to ensure positivity that is
important for nonlinear reconstruction in which a negative weight could cause significant error or even instability. However, the clipping procedure deteriorates accuracy.

It should be noted that the clipping procedure depends on grid information and is independent of flow information \cite{ZhangF2015}. In fact, it is unnecessary to clip
the negative weight for approximating a linear distribution on the grid layout in Fig.\ref{fig:f:no_balence}. The proposed method VWLSQ($n$) calculates linear approximations at first, and postpones the positivity or monotonicity checking to the slope limiter. Therefore, the linear reconstruction will not deteriorate
the reconstruction accuracy by certain grid layouts.

\subsection{Perturbation on triangular grid} \label{sec:Comp2}

Triangular/tetrahedral discretization is a normal choice for complex geometry simulations, and perturbation on the grids is usually occurred due to the complexity of the applications. Therefore, a brief comparison is presented as follow.
\begin{figure}
 \centering
 \includegraphics[height=2.0in]{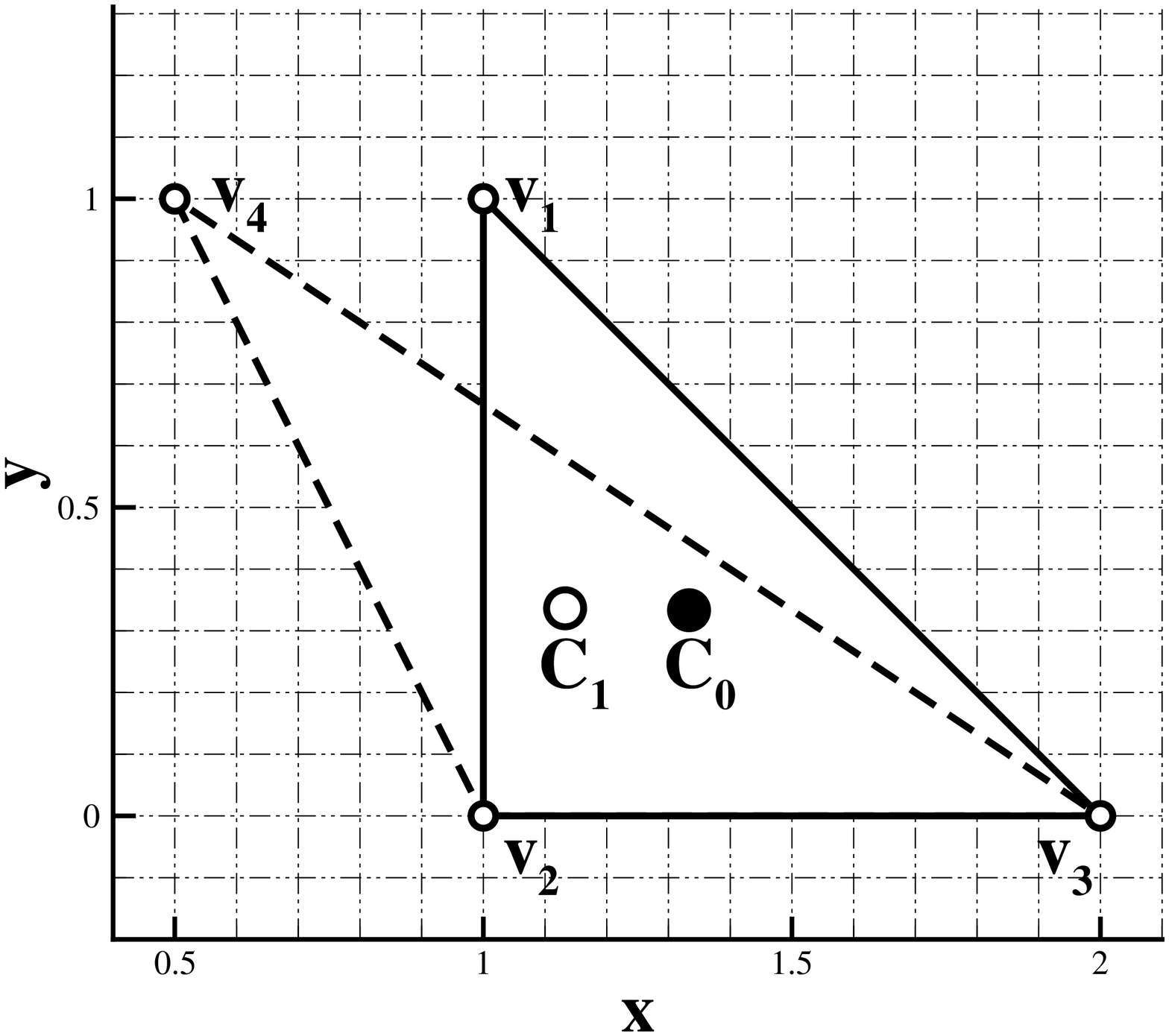}
 \caption{ \label{fig:f:perturbed_tria}
  A perturbed triangular grid }
\end{figure}

In Fig.\ref{fig:f:perturbed_tria}, two triangular cells are given. Without loss of generality, the $x$-direction of the local coordinate is parallel to one of the edges. A function $q=\phi(x,y)=x^2$ is numerically assigned at the grid vertexes. If a perturbation in $x$-direction is assigned
at $v_1$, for example, replace $v_1$ by $v_4$, and then the $x$-direction gradient calculated by GG method at $C_1$ could be written as
\begin{equation}
\overline{\left(\frac{\partial q}{\partial{x}}\right)}_{C_1}^{(\text{GG})} =\frac{\frac{1}{2}l_{34}\left(q_{3}+q_{4}\right)(\mathbf{n}_{{3}{4}})_{x}
-\frac{1}{2}l_{{2}{4}}\left(q_{2}+q_{4}\right)(\mathbf{n}_{42})_{x}}{\Omega_{234}}=3,
\end{equation}
\noindent which is equal to
\begin{equation}
\overline{\left(\frac{\partial q}{\partial{x}}\right)}_{C_0}^{(\text{GG})} =\frac{\frac{1}{2}l_{13}\left(q_{1}+q_{3}\right)(\mathbf{n}_{31})_{x}
-\frac{1}{2}l_{{1}{2}}\left(q_{1}+q_{2}\right)(\mathbf{n}_{{1}{2}})_{x}}{\Omega_{123}}=3.
\end{equation}
These two results indicate that the GG method does not response to the grid perturbation, and thus the accuracy is deteriorated.

On the contrary, the vertex-average procedure for gradient could attain accurate cell-centre gradients. The formulas are
\begin{equation}
\begin{split}
\overline{\left(\frac{\partial q}{\partial{x}}\right)}_{C_0}^{(\text{AVE})} =\frac{\overline{\left(\frac{\partial q}{\partial{x}}\right)}_{v_1}
+\overline{\left(\frac{\partial q}{\partial{x}}\right)}_{v_2}+\overline{\left(\frac{\partial q}{\partial{x}}\right)}_{v_3}}{3}=2x_{C_0}\approx 2.667,  \\
\overline{\left(\frac{\partial q}{\partial{x}}\right)}_{C_1}^{(\text{AVE})} =\frac{\overline{\left(\frac{\partial q}{\partial{x}}\right)}_{v_2}
+\overline{\left(\frac{\partial q}{\partial{x}}\right)}_{v_3}+\overline{\left(\frac{\partial q}{\partial{x}}\right)}_{v_4}}{3}=2x_{C_1}\approx 2.333.
\end{split}
\end{equation}

Here, although the advantage of the vertex-gradient-averaging procedure is not changing the fact that the whole gradient reconstruction scheme
is second-order spatial accuracy, the VWLSQ($n$) could be less sensitive to the grid perturbation.

\subsection{WLSQ scheme on a perturbed high aspect-ratio grid} \label{sec:Comp3}
Inverse distance weight is usually implemented for improving the accuracy of WLSQ schemes. However, WLSQ schemes are sensitive to the small perturbation of computation grids \cite{Diskin2011}.

\begin{figure}
 \centering
 \includegraphics[height=2.5in,angle=-90]{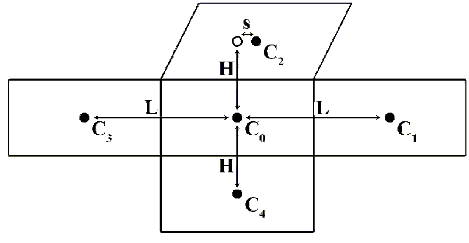}
 \caption{ \label{fig:f:perturbed_quad}
  A perturbed quadrilateral grid }
\end{figure}

In Fig.\ref{fig:f:perturbed_quad}, gradient at cell centre $C_0$ needs to be calculated, and the value at each cell centre is given.
A function $q=\phi(x, y)$ is given on this grid. Cell centres $C_1$, $C_3$, and $C_4$ are perpendicular to $C_0$.
The coordinate of $C_0$ is $\mathbf{x}_{C_0}=(x,y)_{C_0}=(0,0)$, and then the other coordinates are $\mathbf{x}_{C_1}=(L,0)$, $\mathbf{x}_{C_3}=(-L,0)$, and $\mathbf{x}_{C_4}=(0,-H)$.
Here, a perturbation is set at $C_2$, of which the coordinate is $(x,y)_{C_2}=(s,H)$.

If $L\gg H$, and $H\sim s$, it leads to a high aspect-ratio grid with perturbation. If the cell-based WLSQ($1$) scheme is applied on this grid, leaving out
the detail of the derivation, the solution of the least-squares problem is
\begin{equation} 
\begin{aligned}
A^TAX=A^TB
=\left(
  \begin{array}{cccc}
    \omega_{a}L    & \omega_{c}s     & -\omega_{a}L   & 0              \\
    0              & \omega_{c}H     & 0              & -\omega_{b}H   \\
  \end{array}
\right)
\left( \begin{array}{c}  \omega_{a}\Delta q_{{0}{1}} \\ \omega_{c}\Delta q_{{0}{2}}
\\ \omega_{a}\Delta q_{{0}{3}} \\ \omega_{b}\Delta q_{{0}{4}} \\ \end{array} \right)
=\left( \begin{array}{c} \omega_{a}^2L(\Delta q_{{0}{1}}-\Delta q_{{0}{3}})+\omega_{c}^2s\Delta q_{{0}{2}}
 \\ H(\omega_{c}^2\Delta q_{{0}{2}}-\omega_{b}^2\Delta q_{{0}{4}}) \\ \end{array} \right), \\
X=(A^TA)^{-1}A^TB
=\frac{1}{|A^TA|}\left(
  \begin{array}{cc}
    H^2(\omega_{b}^2+\omega_{c}^2)   & -\omega_{c}sH       \\
    -\omega_{c}sH           &  2\omega_{a}^2L^2+\omega_{c}^2s^2  \\
  \end{array}
\right)
\left( \begin{array}{c} \omega_{a}^2L(\Delta q_{{0}{1}}-\Delta q_{{0}{3}})+\omega_{c}^2s\Delta q_{{0}{2}}
 \\ H(\omega_{c}^2\Delta q_{{0}{2}}-\omega_{b}^2\Delta q_{{0}{4}}) \\ \end{array} \right).
\end{aligned}
\end{equation}

\noindent Here, $\omega_a=\omega_1=\omega_3=\frac{1}{L}$, $\omega_b=\omega_4=\frac{1}{H}$, $\omega_c=\omega_2=\frac{1}{\sqrt{H^2+s^2}}$,
thus $\omega_b\sim \omega_c\gg \omega_a$. As a result, the gradient
 is nearly independent of the variables at $C_1$ and $C_3$.
It means slightly perturbation on high aspect-ratio grid will severely deteriorate the accuracy of cell-based WLSQ schemes, especially
on the boundary layer grid on which variables are changed violently in the vertical direction.

On the other hand, the vertex-based VWLSQ($n$) scheme is less sensitive to such weighting effect because its weights are approximately in the same magnitude, and thus the contribution
of each stencil cell will be utilized. Therefore, the proposed method
is expected to be more accurate and stable on perturbed grids. In the following numerical cases, the VWLSQ($1$) is showing better performance on high aspect-ratio grids compared with cell-based WLSQ schemes.

\subsection{Curvature grid and geometrical monotonicity}
\begin{figure}
 \centering
 \includegraphics[width=1.2in,angle=-90]{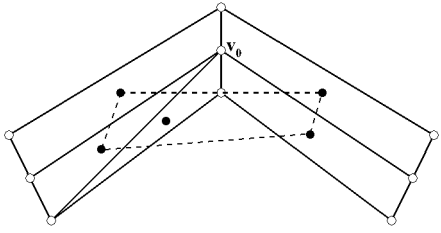}
 \caption{ \label{fig:f:curved_grid}
  A curved hybrid grid }
\end{figure}

The geometrical monotonicity problem was investigated, for which a hybrid formula of GG and WLSQ(G), GLSQ \cite{Shima2010,Shima2013}, was proposed. In the research of
GLSQ, GG method had been used to fix the geometrical monotonicity violation induced by WLSQ scheme. For a vertex-based gradient reconstruction scheme, for instance the PL-GG scheme, is also suffering this monotonicity violation. As shown in Fig.\ref{fig:f:curved_grid}, flow information at the vertex
$v_0$ is expected to be calculated by the neighbouring cell-centre variables. However, $v_0$, which is outside the dashed-line zone,
 leads to an extrapolation, instead of an interpolation, in the reconstruction. Therefore, in order to guarantee monotonicity, clipping procedure is used, which leads to the PL-CLIP-GG scheme.

As mentioned in section \ref{sec:clip}, it is not always necessary to invoke the clipping procedure, even negative weights are produced.
The presented VWLSQ($n$) scheme which is without clipping procedure does not guarantee monotonicity. However, the numerical results will prove the scheme remains stable on many cases, on
which the traditional cell-based WLSQ scheme is unstable or even blow-up.

\section{Numerical results} \label{sec:tests}

 \subsection{Approximation of analytical function on perturbed grids}

\begin{figure}
 \centering
 \subfigure[Type \emph{I}: quadrilateral grid] %
  {\label{fig:f:I}\includegraphics[width=1.8in]{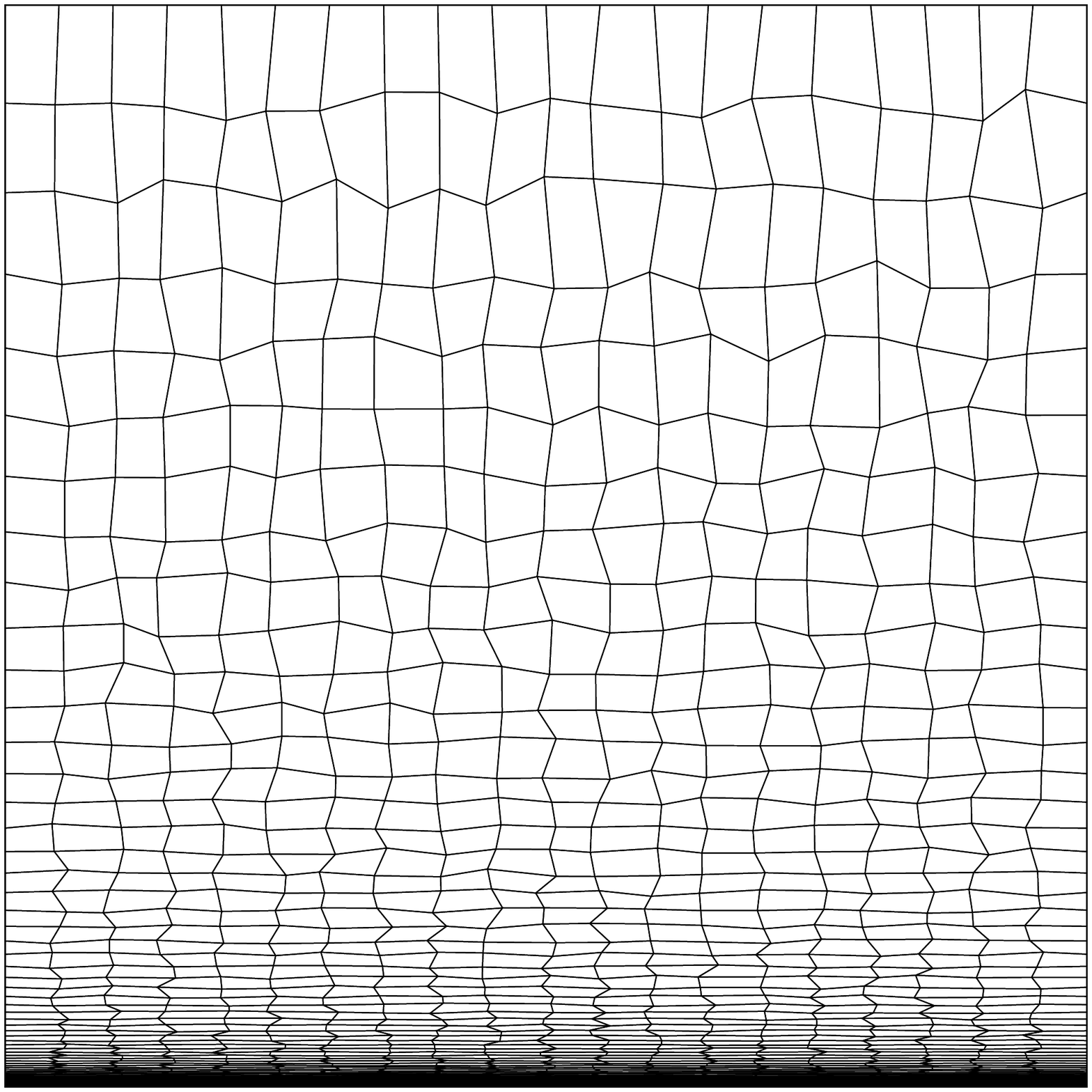}}
 \subfigure[Type \emph{II}: regular triangular grid] %
  {\label{fig:f:II}\includegraphics[width=1.8in]{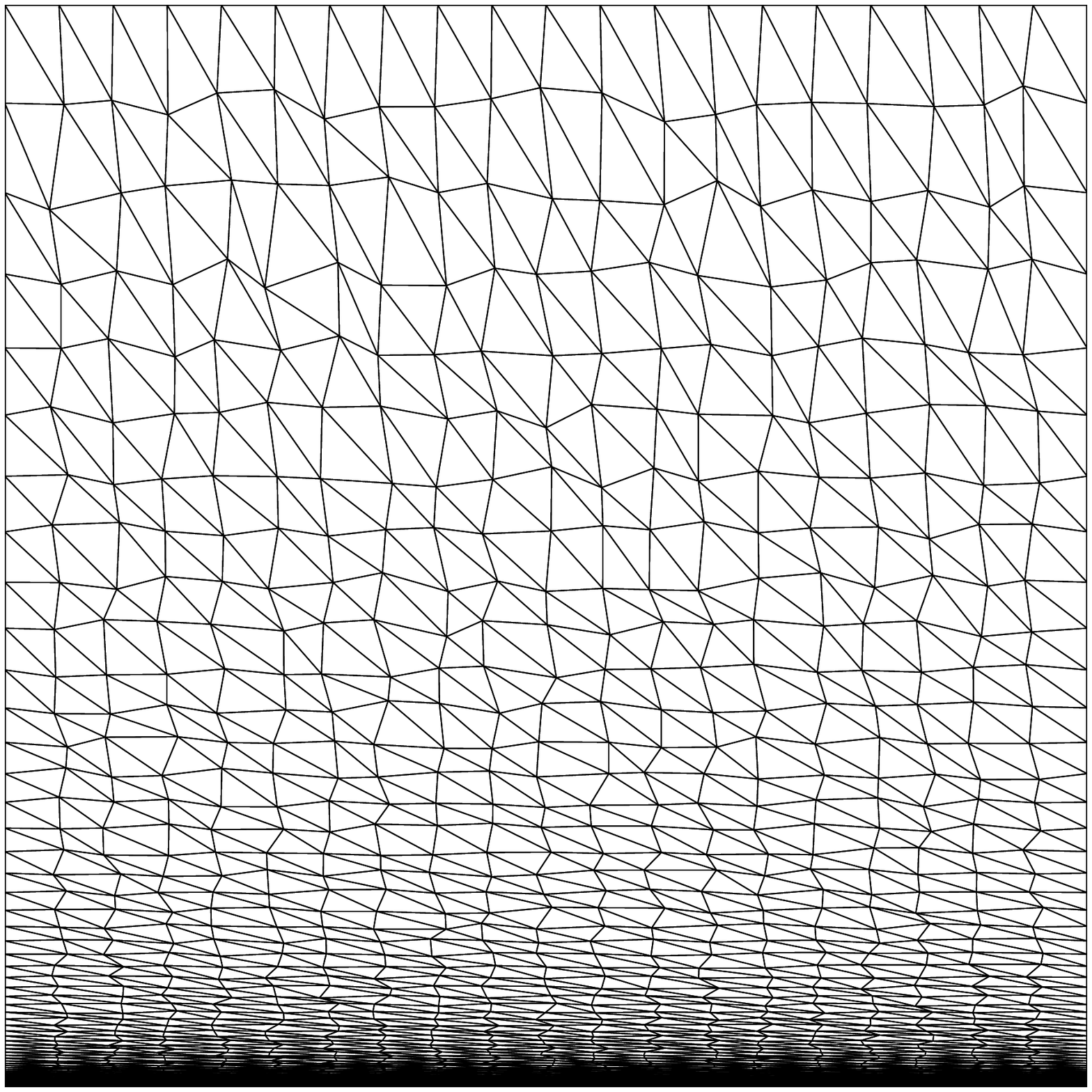}}\\
   \subfigure[Type \emph{III}: irregular triangular grid] %
  {\label{fig:f:III}\includegraphics[width=1.8in]{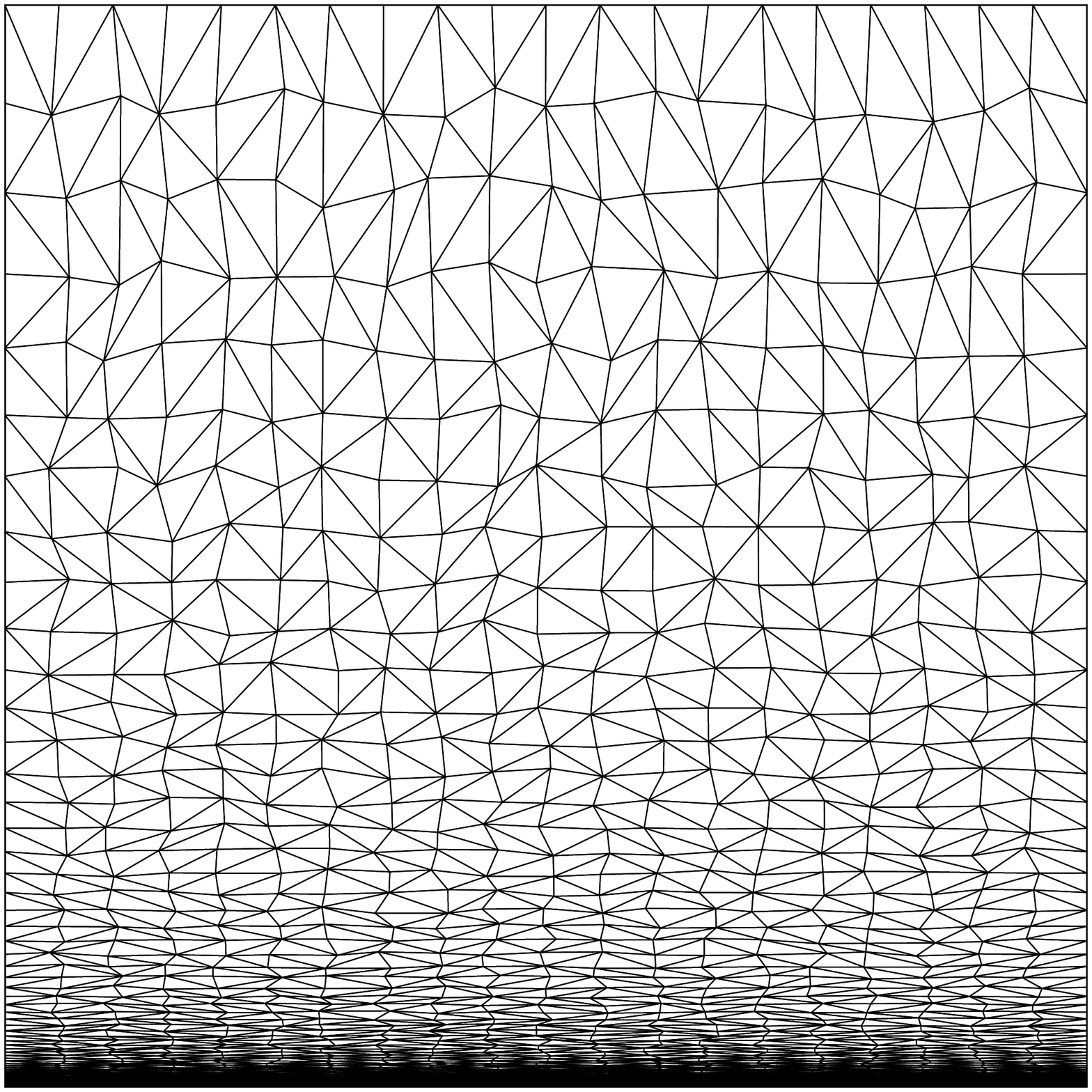}}
 \subfigure[Type \emph{IV}: random hybrid grid] %
  {\label{fig:f:IV}\includegraphics[width=1.8in]{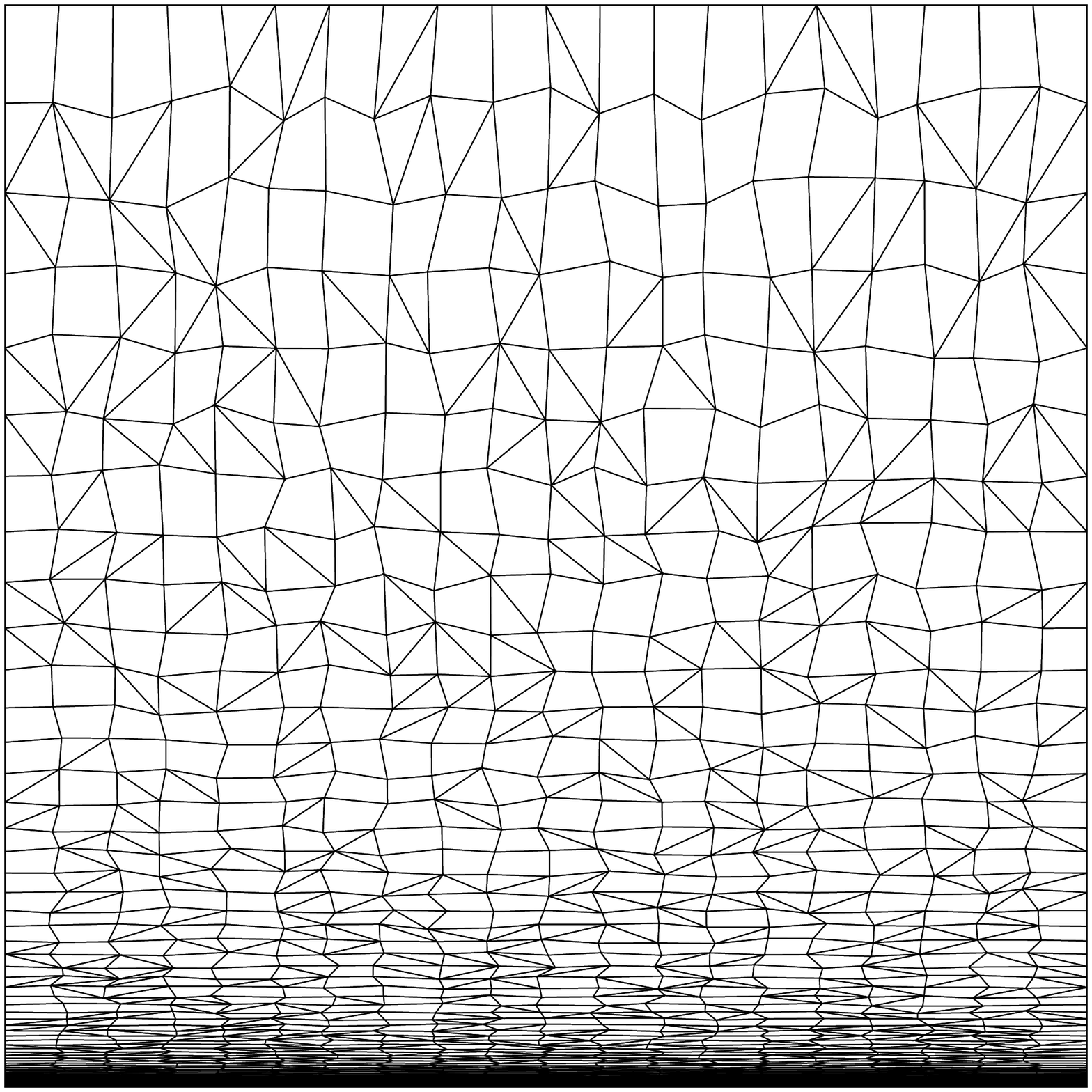}}
 \caption{ \label{fig:f:grids}
  Perturbed test grids}
\end{figure}

In this section, perturbed grids are used to discretize a rectangular zone, as shown in Fig.\ref{fig:f:grids}. The test grids are devised to bear the conditions of high aspect-ratio and skewness,
and then some basic properties of the proposed method could be shown.

The zone is $1\times 1$ rectangular geometry and containing $21 \text{(horizontal)} \times 101 \text{(veritical)}$ vertexes.
The two triangular grids is made by bisecting the quadrilateral grid regularly and randomly, respectively.
Width of each cell is identical, that is $w = 1/20 = 0.05$, and the height of the first layer at the bottom is $h = 7.25719 \times 10^{-6}$, and thus the maximum aspect-ratio is
$6890$ for the quadrilateral cells and $9186$ ($6890 \times 4/3$) for the triangular cells. Cell height grows from the bottom to the top in a spacing rate of $1.1$.
The random perturbation of grid vertexes is in range of $(-0.1,0.1) \times w$ in horizontal and $(-0.1,0.1) \times h$ in vertical.

To compare the accuracy of gradient reconstruction methods, a simple analytical function $q=y^2$ is numerically set at all the vertexes.
Therefore, the accurate gradients are known. The function $\nabla q_y$ is the vertical directional derivative of $q=2y$.
Here, VWLSQ($0$) and VWLSQ($1$) are showing nearly identical results, and thus the results of VWLSQ($1$) scheme are omitted.

\begin{figure}
 \centering
 \includegraphics[width=2.2in]{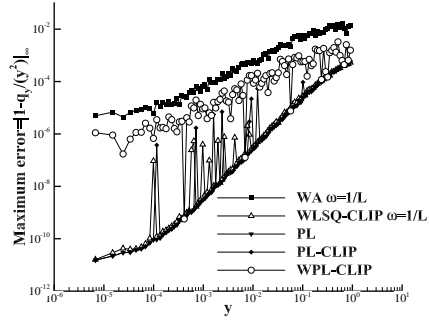}
 \caption{ \label{fig:f:NA_result_grid4}
 Vertex value accuracy of NA methods on perturbed grid \emph{IV}}
\end{figure}

Firstly, the vertex values reconstruction errors of NA methods on grid $IV$ are shown in Fig.\ref{fig:f:NA_result_grid4} to explain some of
the basic characters of the schemes. It is obvious that the clipping procedure causes significant errors.
It should be noted that the weighted PL scheme, WPL, with using clipping procedure, produces significant error which is approximately equal to the WA scheme.
On the other hand, PL scheme without clipping shows better and stable accuracy.
Therefore, the clipping procedure is unnecessary in certain condition, which is not always detectable preliminary.

\begin{figure}
 \centering
 \subfigure[Grid \emph{I}] %
  {\label{fig:f:G1}\includegraphics[width=2.2in]{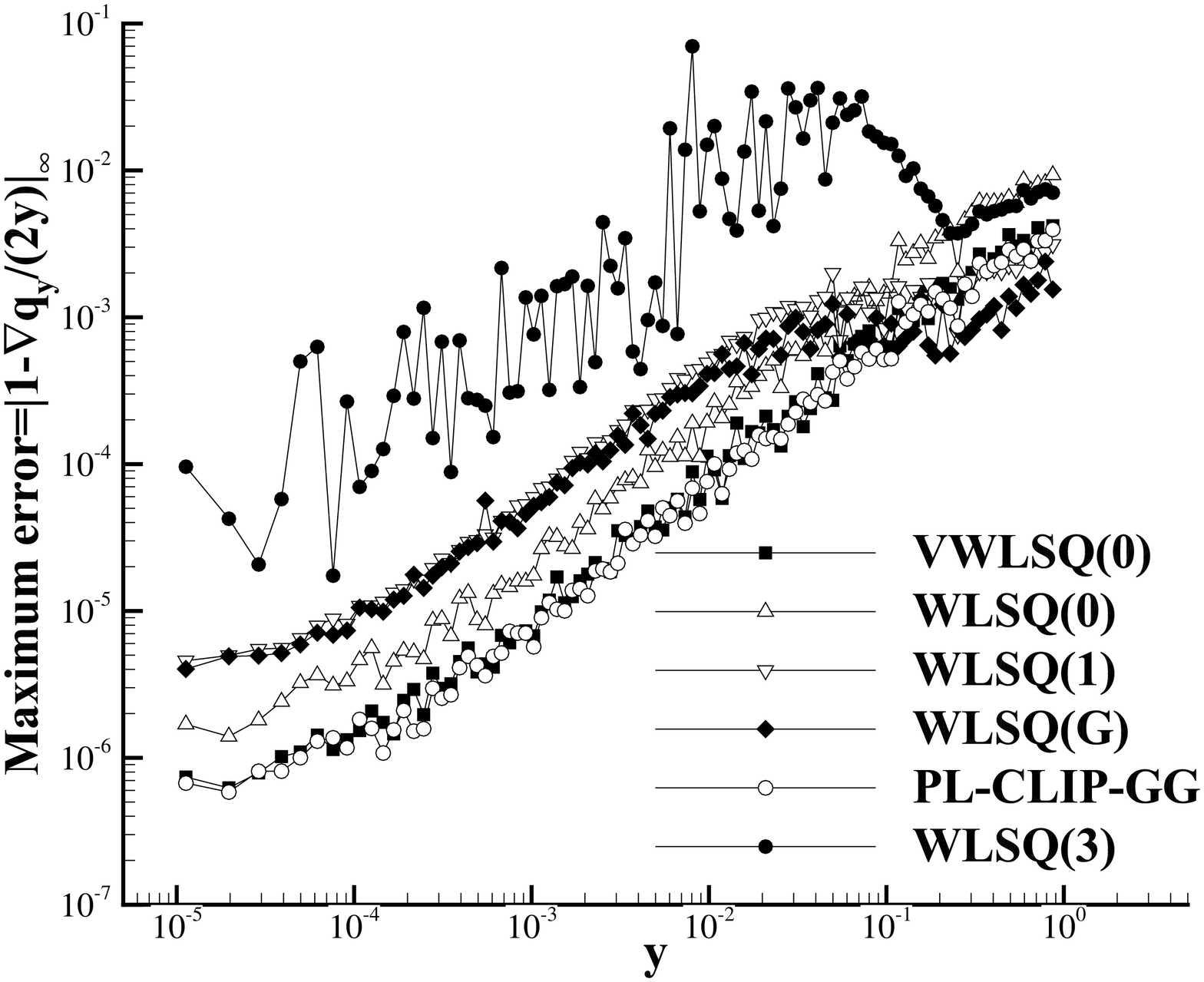}}
 \subfigure[Grid \emph{II}] %
  {\label{fig:f:GII}\includegraphics[width=2.2in]{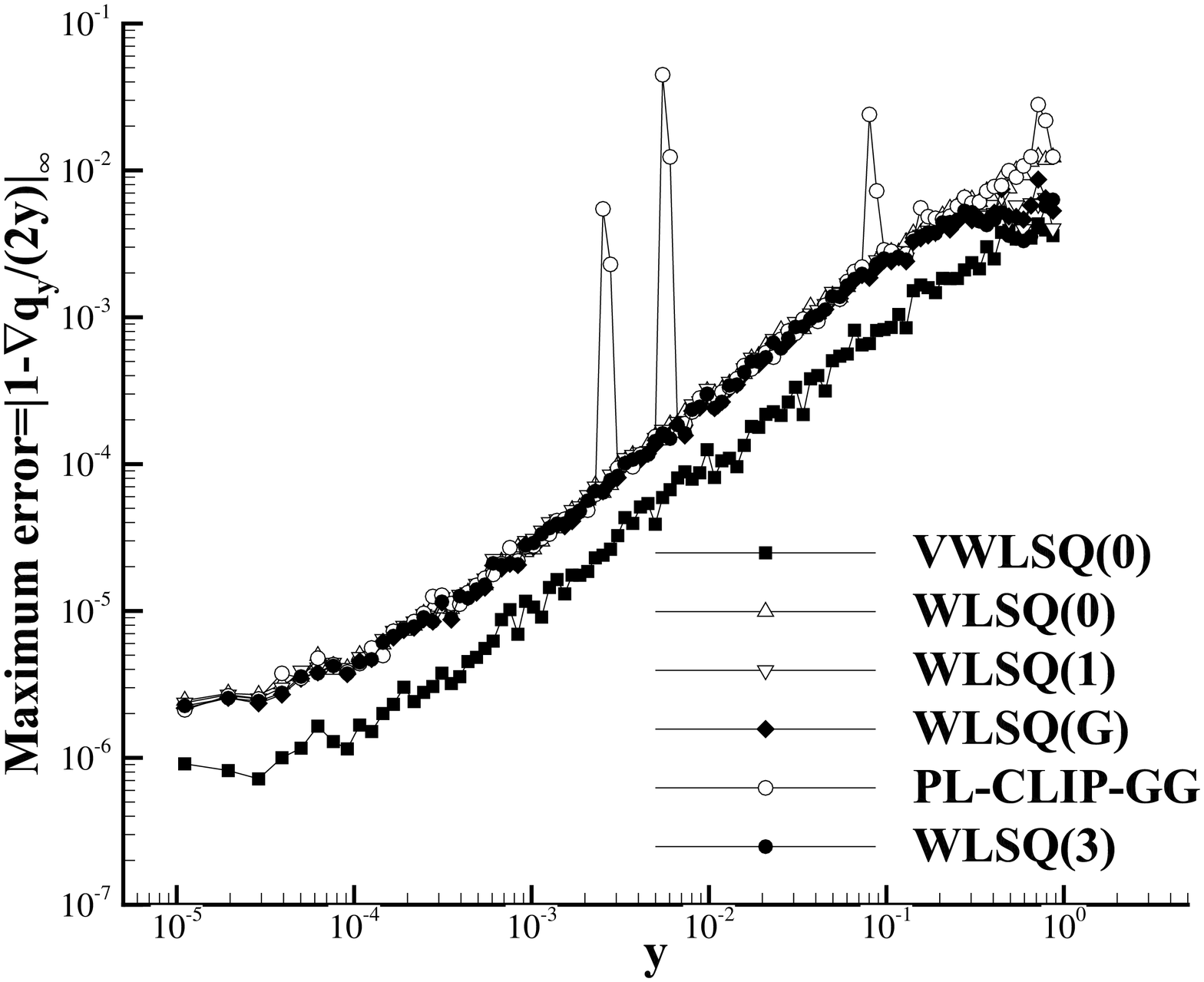}}
   \subfigure[Grid \emph{III}] %
  {\label{fig:f:GIII}\includegraphics[width=2.2in]{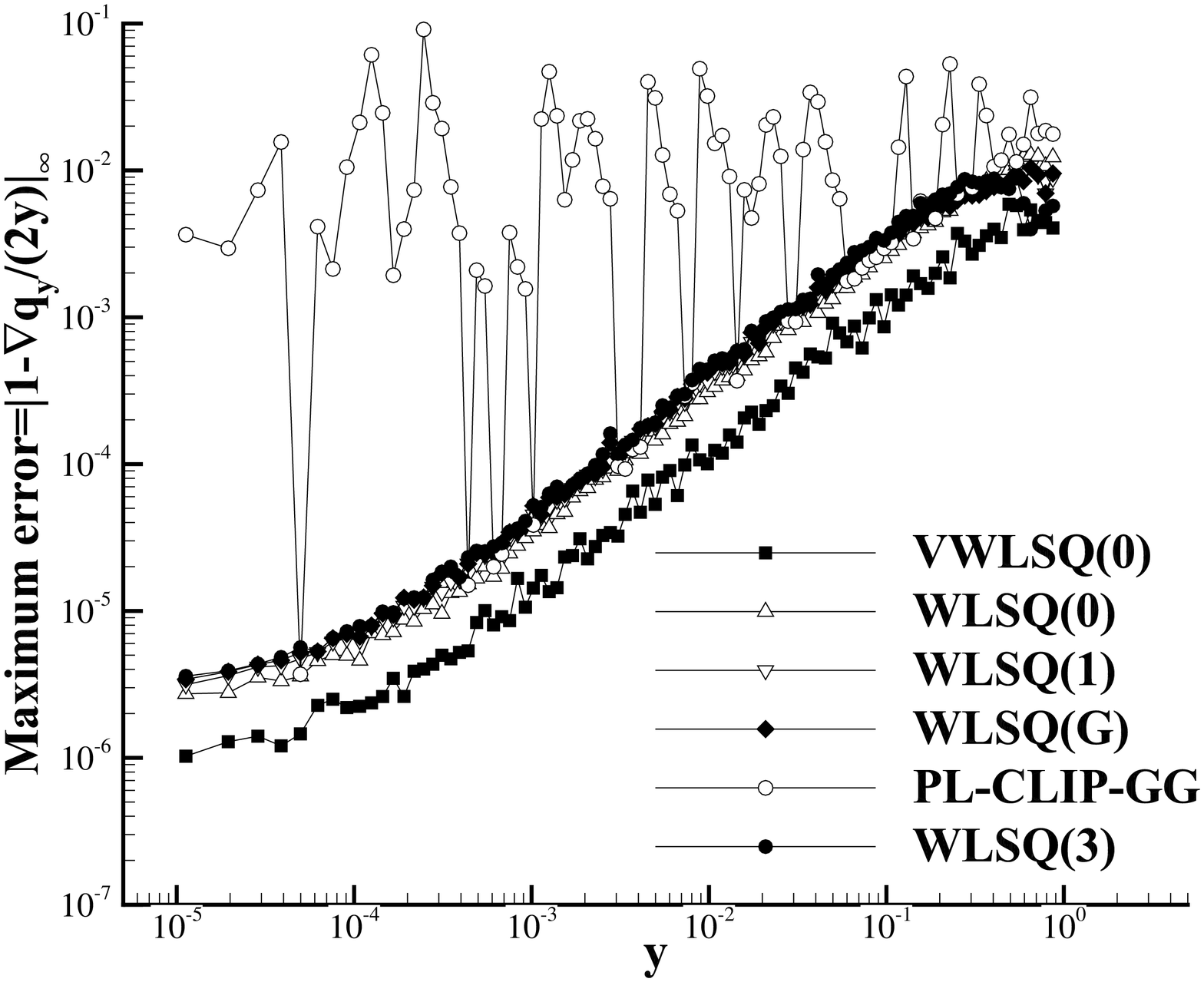}}
 \subfigure[Grid \emph{IV}] %
  {\label{fig:f:GIV}\includegraphics[width=2.2in]{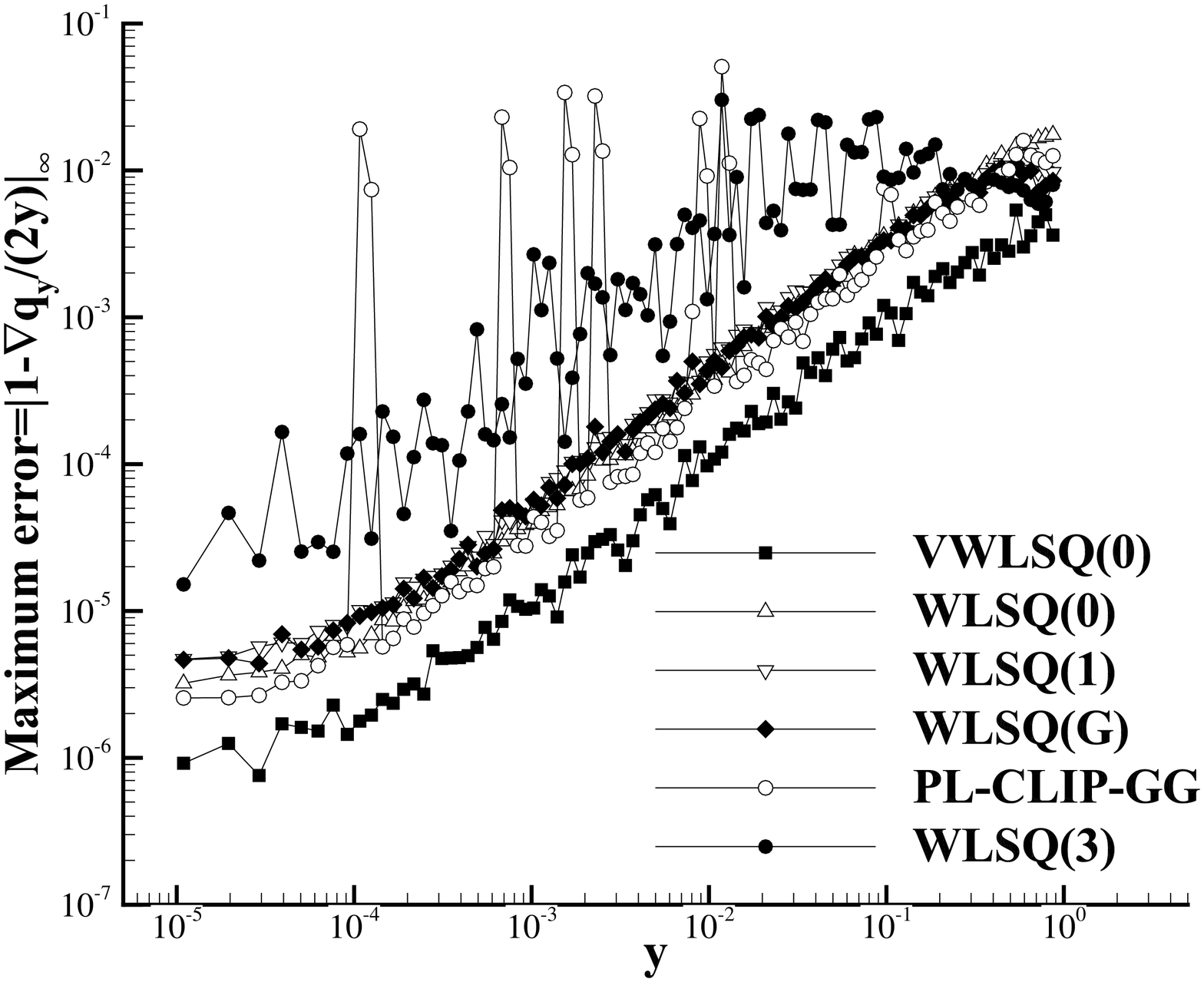}}
 \caption{ \label{fig:f:Grid_results}
  Approximation errors on perturbed test grids}
\end{figure}

Subsequently, the cell-centre gradient reconstruction errors are shown in Fig.\ref{fig:f:Grid_results}.
The results of NA procedures showed that WA and WPL-CLIP scheme are less accurate on the given grid, and thus only the PL scheme is used in the following
comparison.
 Again, the clipping procedure causes oscillate errors, which is expectable after the former results. The cell-based WLSQ(3) scheme shows larger errors in quadrilateral cells, and thus the errors are more significant on
grid $I$ and $IV$. This is a proof of the conclusion in section \ref{sec:Comp3}. It should be noted that, for cell-based WLSQ scheme, weighting improves the results in low aspect-ratio cells. In general, VWLSQ scheme is showing the best result in all the grid conditions.

 \subsection{Subsonic inviscid incompressible flow around a cylinder}
The former test case only shows gradient approximation at cell centres, and thus it is not sufficient to explain the convergence and stability of each scheme.
Therefore, this test case is given.
The flow condition of the presented test is that inviscid flow of which the uniform inflow Mach number is set as 0.3 goes around a circular cylinder.
Flows in this given Mach number is approximately incompressible, and thus there is not drag in the cylinder theoretically. Furthermore,
the entropy should not increase or decrease in all the flow field. Therefore, any non-zero drag of the cylinder, or entropy variation in flow field, indicate numerical error of numerical schemes.
Because there is not discontinuity in the flow field, slop limiters are not used, and thus the gradient reconstruction schemes could be investigated solely.

\begin{figure}
 \centering
 \subfigure[Quadrilateral grid] %
  {\label{fig:f:meshquad}\includegraphics[width=2.0in]{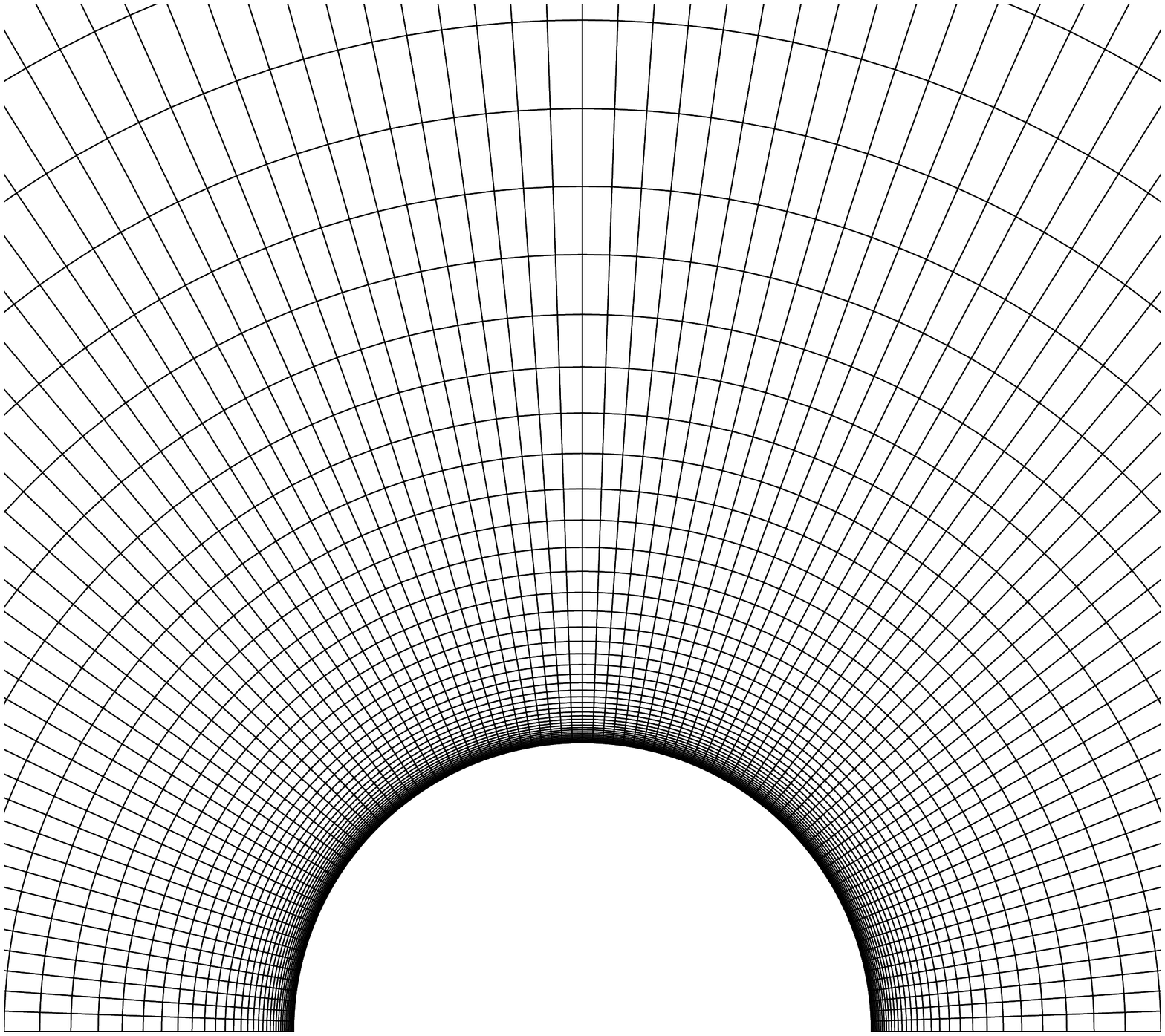}}
 \subfigure[Triangular grid] %
  {\label{fig:f:meshtria}\includegraphics[width=2.0in]{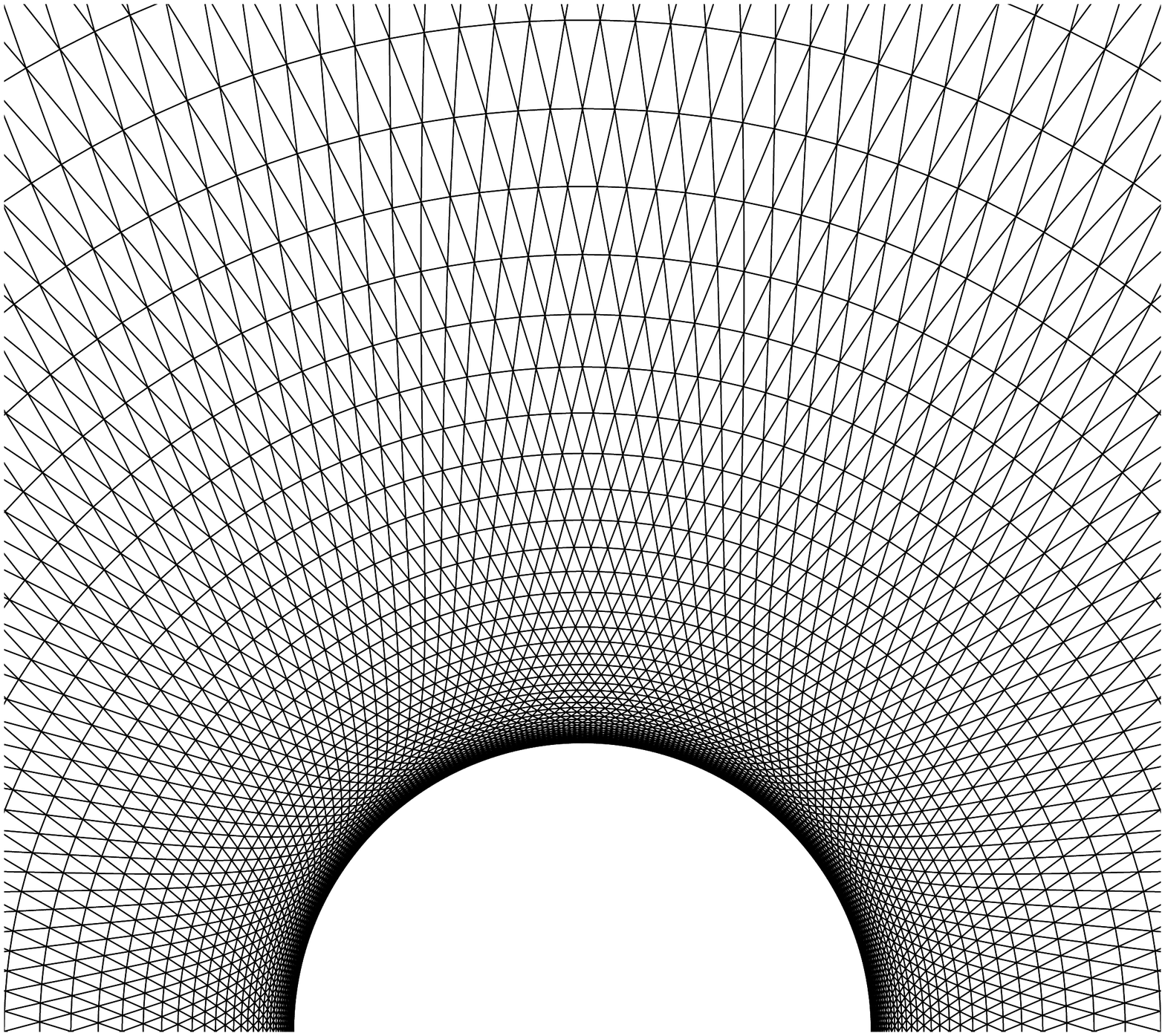}}
 \caption{ \label{fig:f:Inv_Grids}
  Grids for the simulations of inviscid flow around a cylinder (close-up view)}
\end{figure}

The diameter of the cylinder is $D=1$, and the diameter of the circular outer boundary is $40D$.
Two quadrilateral discretizations are defined. The coarse one is $180(\text{Circumferential})\times60(\text{radial})$ and the fine one is $270\times90$, of which the first layer cells are $0.02D$ and $0.01D$ respectively.
Correspondingly, two triangular grids are defined. From the inner slip wall to
the outer non-reflecting boundary, the layers and cell heights
of triangular grids are as same as those of the quadrilateral grids, but the cells of each layer of triangular grids are double those of the quadrilateral grids. All the triangular cells are isosceles triangle and the vertex angles point to the outer boundary.
Coarse quadrilateral and triangular grids are shown in Fig.\ref{fig:f:Inv_Grids}.

\begin{figure}
 \centering
 \subfigure[Pressure] %
  {\label{fig:f:cylinder_p}\includegraphics[width=2.4in]{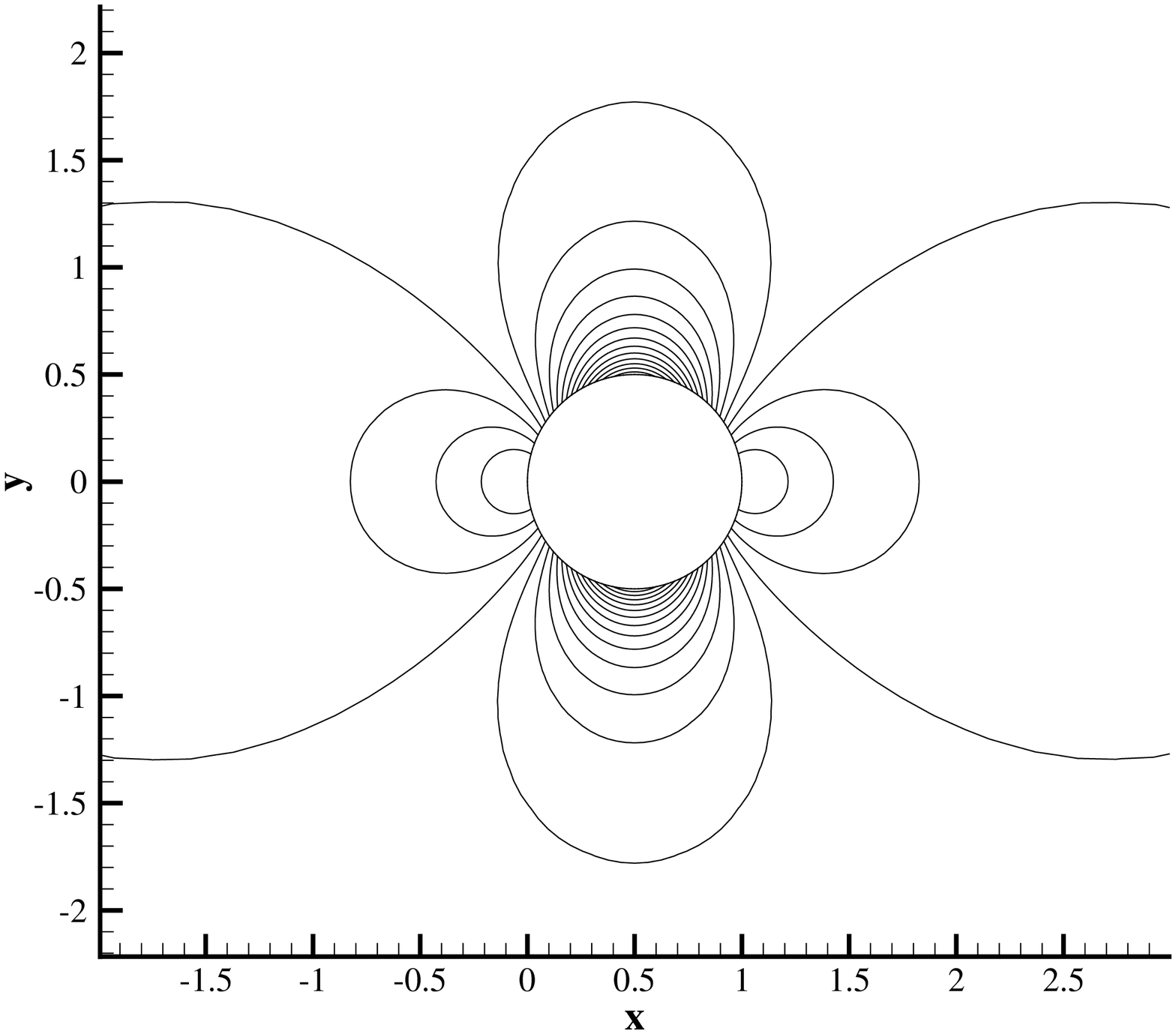}}
 \subfigure[$y$-velocity] %
  {\label{fig:f:cylinder_v}\includegraphics[width=2.4in]{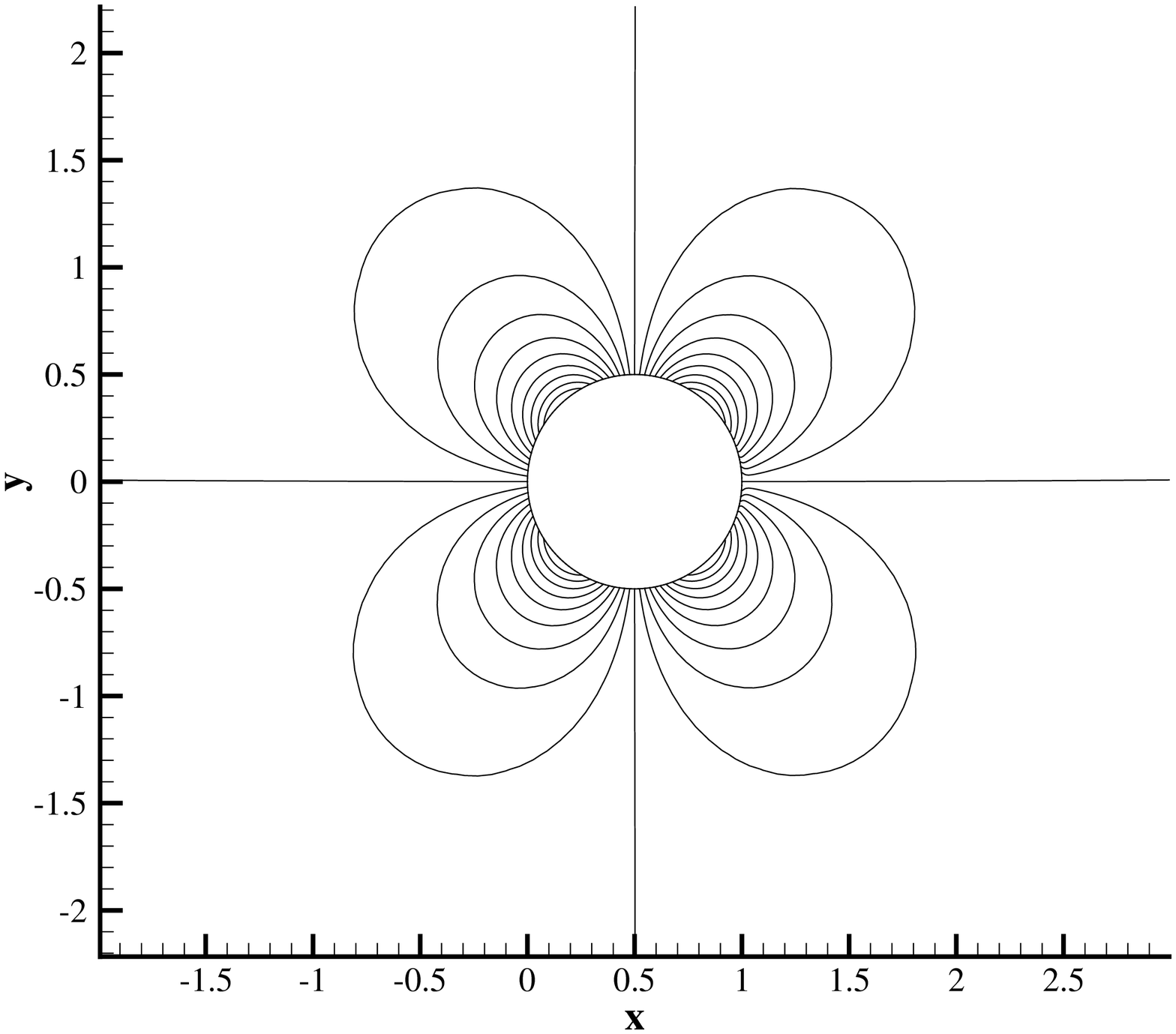}}
 \caption{ \label{fig:f:Inv_results}
  Flow contours of the inviscid flow around a circular cylinder problem}
\end{figure}

\begin{figure}
 \centering
 \subfigure[Quadrilateral grid] %
  {\label{fig:f:quad_res}\includegraphics[width=2.4in]{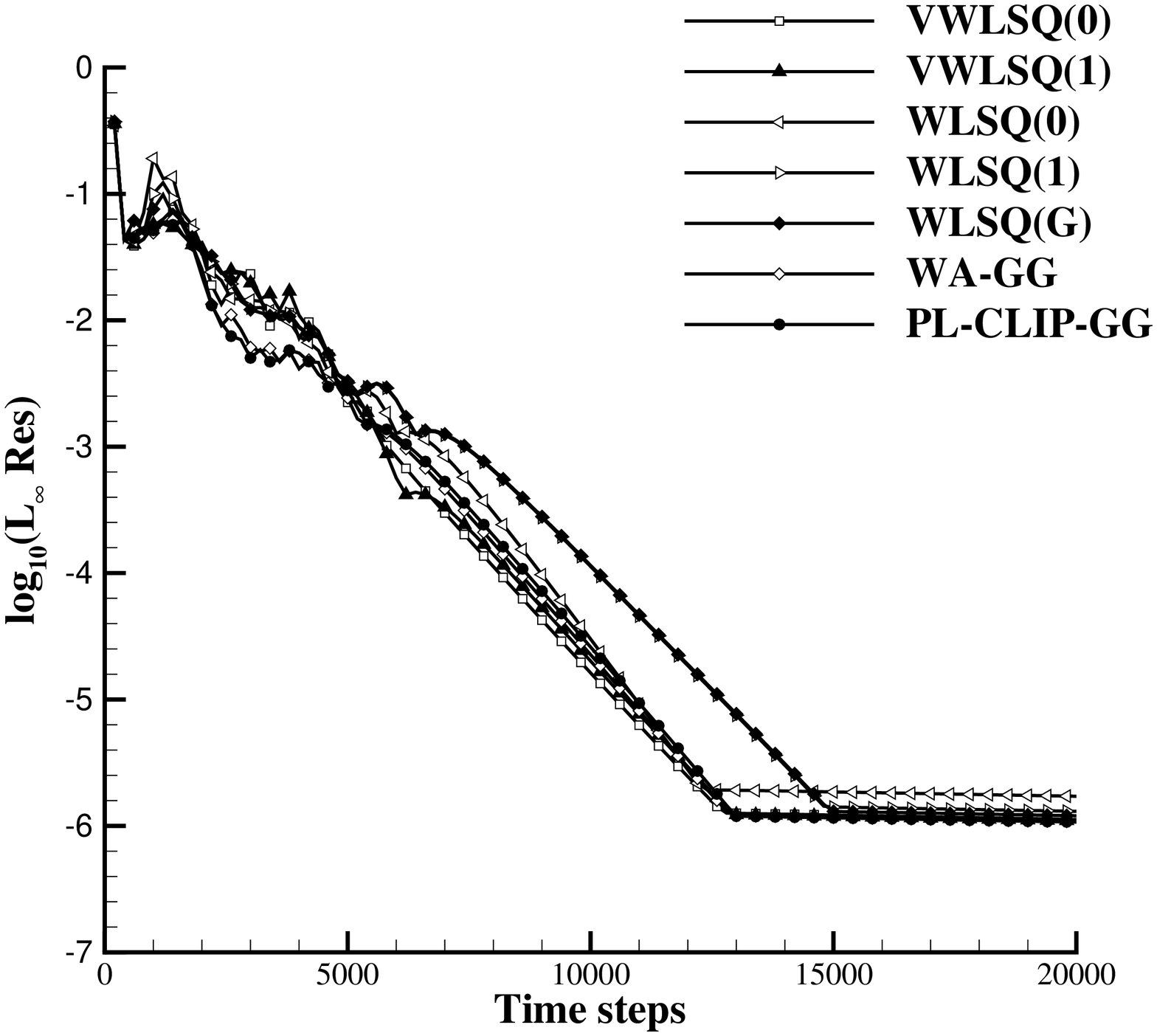}}
 \subfigure[Triangular grid] %
  {\label{fig:f:tria_res}\includegraphics[width=2.4in]{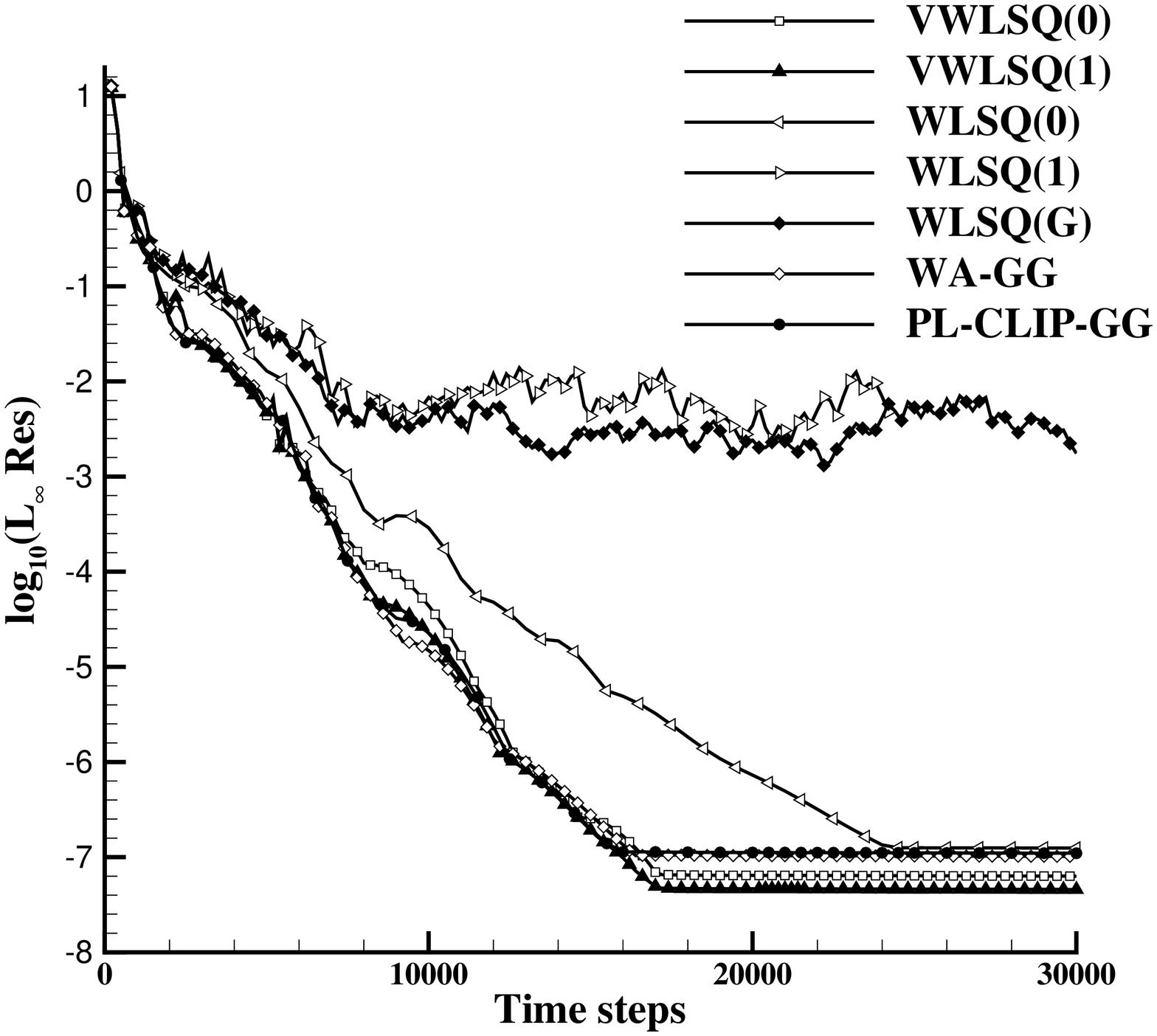}}
 \caption{ \label{fig:f:Inv_res}
 Computational residuals of subsonic inviscid flow around a cylinder on coarse grids}
\end{figure}

The flow condition is steady state, and thus the temporal accuracy is not required here. The commonly used implicit scheme LU-SGS \cite{YOON1988} is implemented for the solution of
the semi-discrete equations, with $CFL=100$ defined.
The HLLC scheme \cite{Toro1994} is used for the calculation of convective fluxes.
Flow contours of VWLSQ(1) scheme are shown in Fig.\ref{fig:f:Inv_results}, as an example. Due to the property of inviscid incompressible flow, the contours are symmetrical. Different schemes
show little difference in contour, but the convergence histories in Fig.\ref{fig:f:Inv_res} are significantly different.

Because the schemes on coarse grids or fine grids are showing similar convergence pattern, relatively, only the results on coarse grids are given in the figures.
On quadrilateral grid, the convergence of each scheme is similar except the WLSQ(1) and WLSQ(G) cost more computation steps to be convergent.
The vertex-based schemes show significant advantage compared with the cell-based WLSQ scheme on triangular grids. Thereinto, WLSQ(1) and WLSQ(G)
are unable to converge and WLSQ(0) costs more computation steps compared with vertex-based schemes. Two VWLSQ schemes show a little advantage in converged residual among
the vertex-based schemes.

Then the accuracy is compared.
As aforementioned, the drag of the cylinder in this computation condition should be zero in theory. Here, the drag of each scheme on each grid is shown in Table
 \ref{tab:Cd}. While the grid is refined, the darg, $i.e.$ the error, is reduced, with using any of the reconstruction method. The errors of VWLSQ(1) scheme is the smallest among the vertex-based schemes.
The cell-based schemes are showing larger error in most of the cases compared with the VWLSQ(1) scheme, except several exceptions.
The WLSQ(G) scheme is showing the best accuracy on quadrilateral grids, and the WLSQ(1) scheme is showing the best accuracy on the fine triangular grid.
However, their convergence is less admissible which has been shown.

\begin{table}
  \centering
   \caption{Numerical drag coefficients $abs(C_d)$ of the cylinder} \label{tab:Cd}
  \begin{tabular}{lccccc}
    \hline
    & quadrilateral  & triangular &quadrilateral(fine) &  triangular(fine)
   \\\hline
   VWLSQ($0$)   &$2.02\times 10^{-3}$  &$4.01\times 10^{-4}$   &$6.64\times 10^{-4}$  &$1.66\times 10^{-4}$\\
   VWLSQ($1$)   &$3.40\times 10^{-4}$  &$4.26\times 10^{-5}$   &$1.09\times 10^{-4}$  &$1.88\times 10^{-5}$\\
   WLSQ($0$)    &$4.42\times 10^{-3}$  &$1.80\times 10^{-4}$   &$1.49\times 10^{-3}$  &$1.04\times 10^{-4}$\\
   WLSQ($1$)    &$1.18\times 10^{-3}$  &$1.04\times 10^{-4}$   &$4.05\times 10^{-4}$  &$5.47\times 10^{-6}$\\
   WLSQ(G)      &$5.36\times 10^{-5}$ &$3.70\times 10^{-4}$   &$1.04\times 10^{-5}$ &$8.21\times 10^{-5}$\\
   WA-GG        &$5.75\times 10^{-4} $ &   $1.19\times 10^{-4}$ &  $2.00\times 10^{-4}$      &   $7.50\times 10^{-5}$   \\
   PL-CLIP-GG   &$4.21\times 10^{-4}$  &$9.88\times 10^{-5}$   &$1.45\times 10^{-4}$  &$6.75\times 10^{-5}$\\
   \hline
  \end{tabular}
\end{table}

\begin{figure}
 \centering
 \subfigure[VWLSQ(1) (coarse)] %
  {\label{fig:f:s_VWLSQ_t_sp}\includegraphics[width=2.4in,angle=-90]{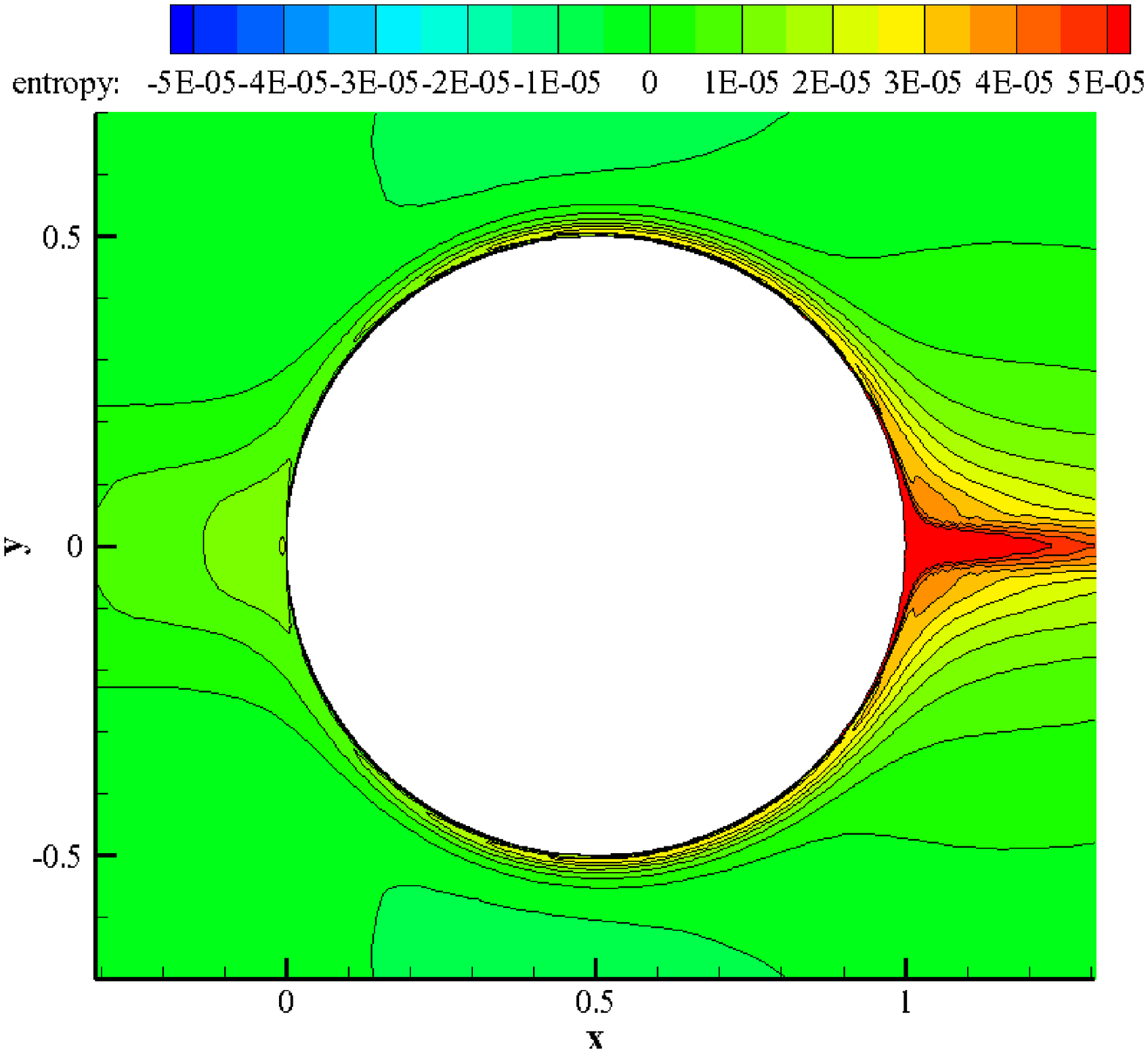}}
 \subfigure[PL-CLIP-GG (coarse)] %
  {\label{fig:f:s_PL_t_sp}\includegraphics[width=2.4in,angle=-90]{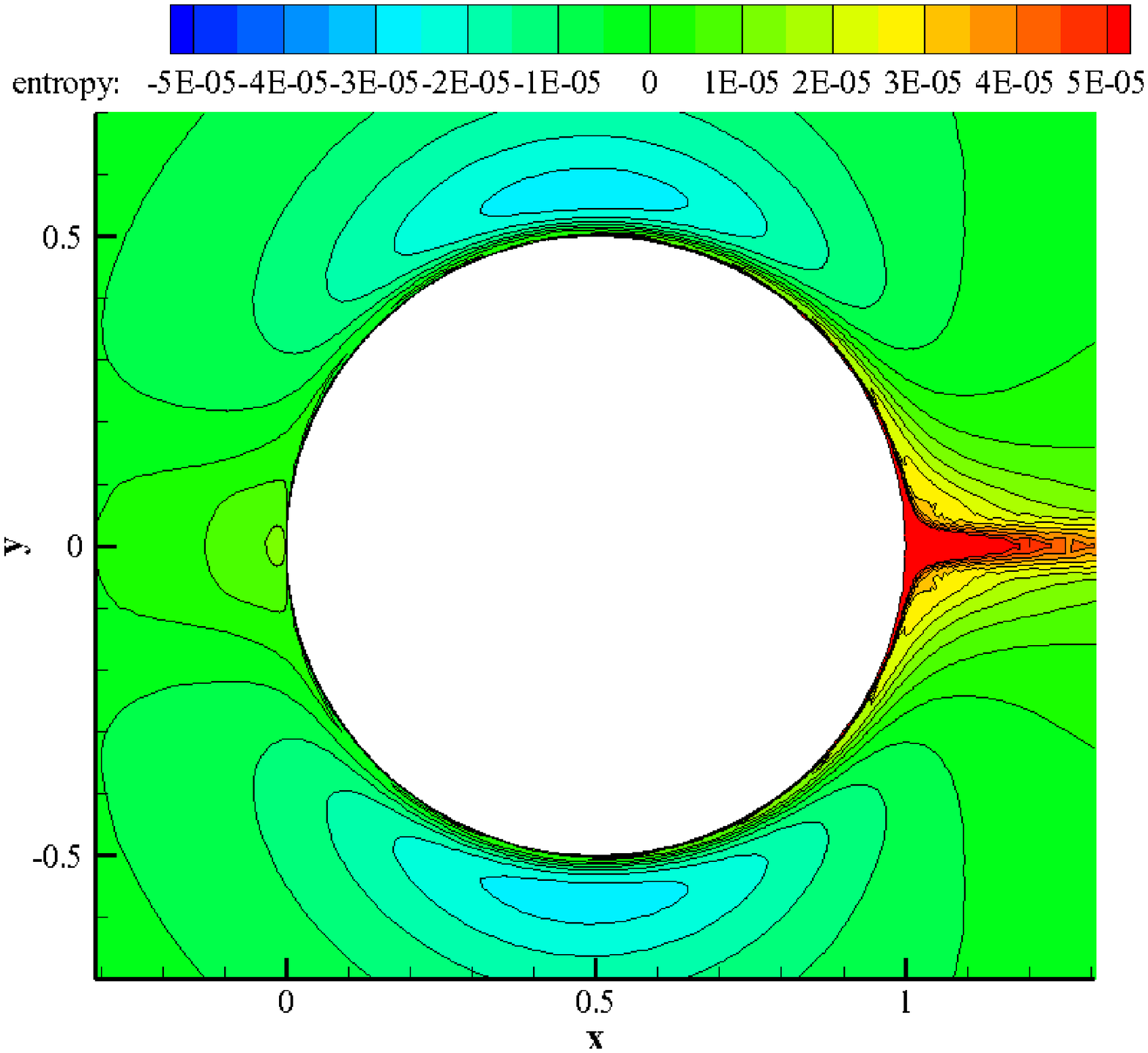}}
   \subfigure[VWLSQ(1) (fine)] %
  {\label{fig:f:s_VWLSQ_t_fi}\includegraphics[width=2.4in,angle=-90]{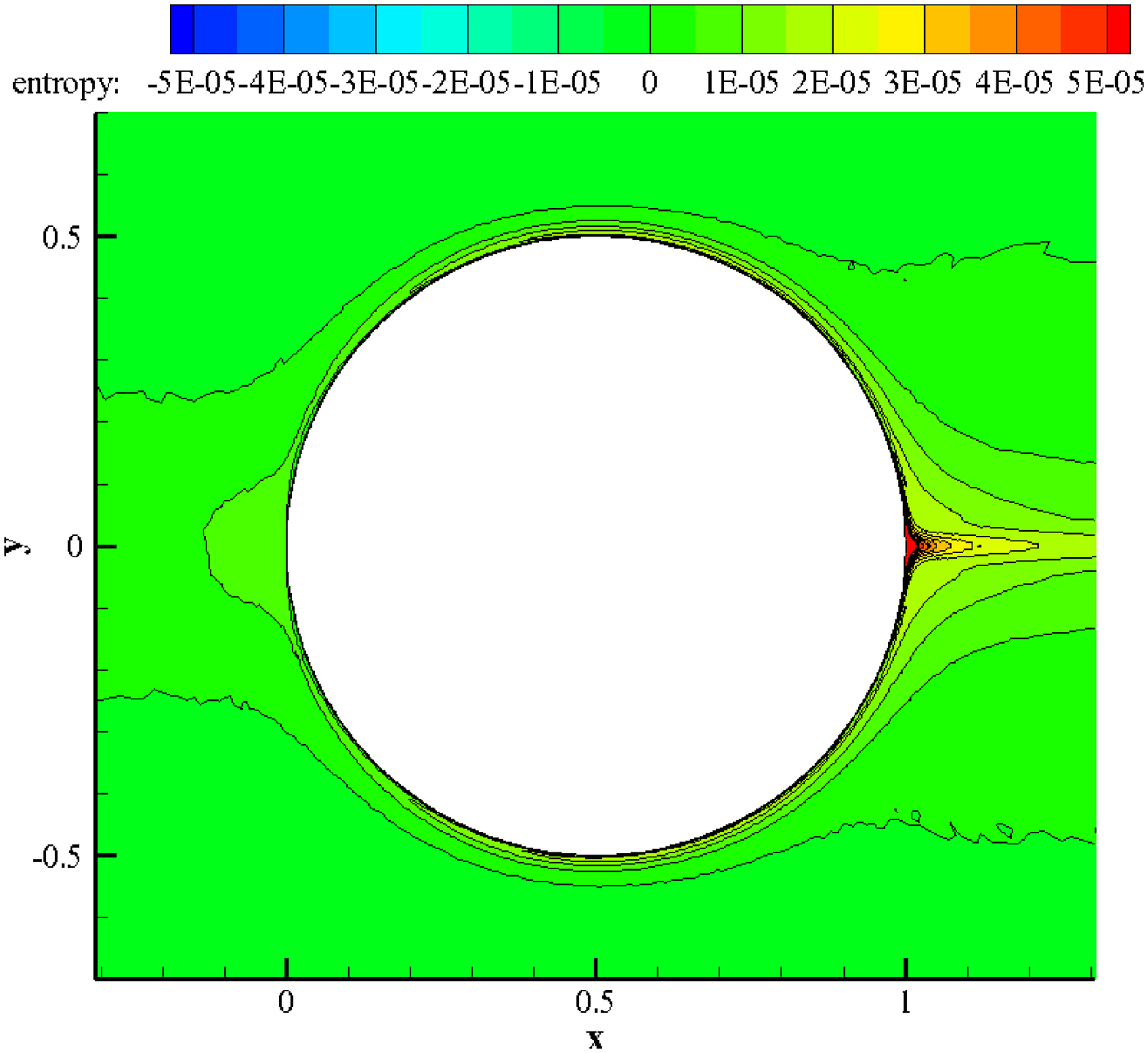}}
 \subfigure[WLSQ(1) (fine)] %
  {\label{fig:f:s_WLSQ_t_fi}\includegraphics[width=2.4in,angle=-90]{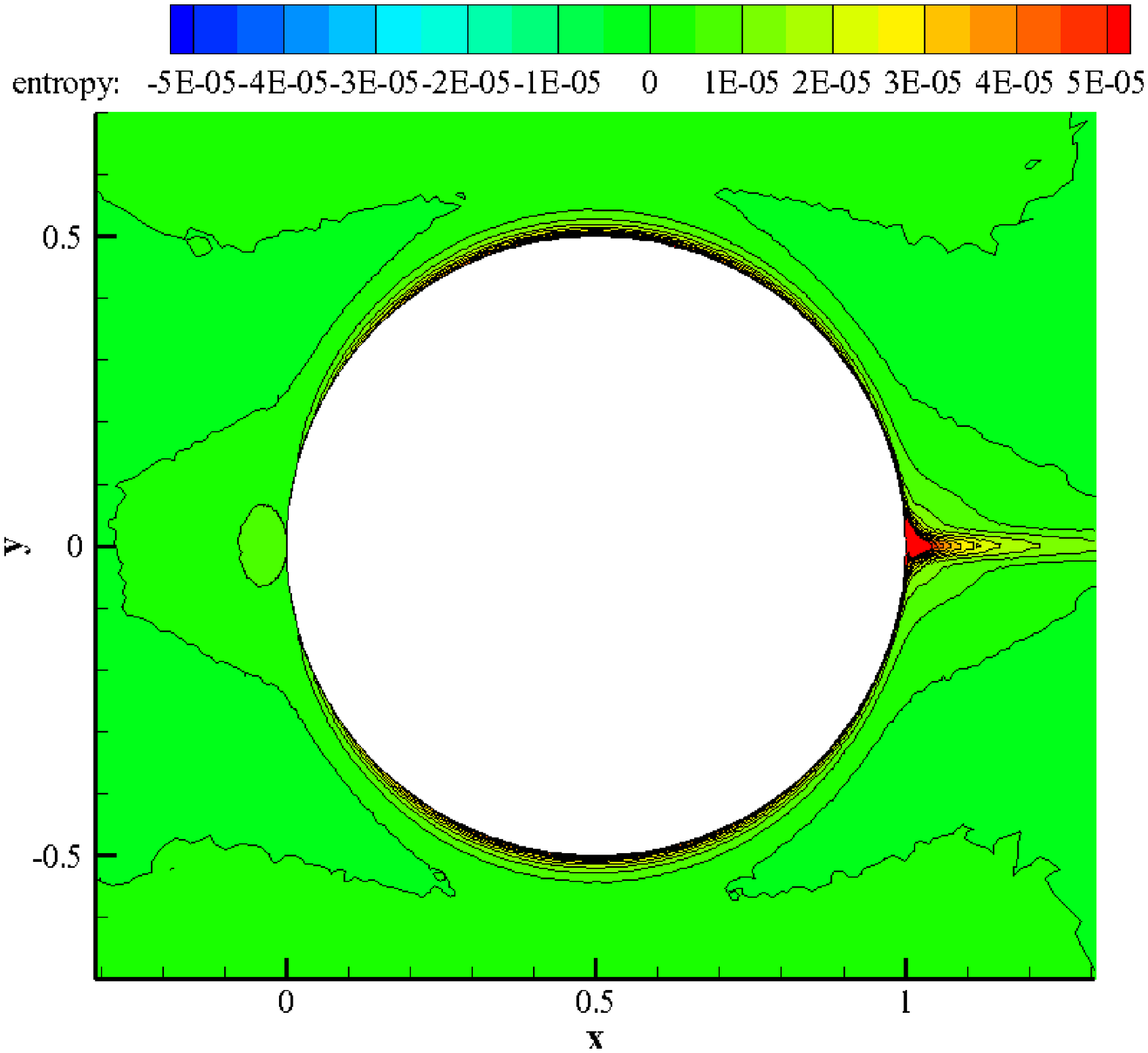}}
 \caption{ \label{fig:f:Inv_s}
  Entropy distribution on triangular grids}
\end{figure}

Besides the integral error, absolute drag of cylinder, the distributed error, entropy, which is defined as $\ln[(p/\rho^\gamma)/(p_{\infty}/\rho_{\infty}^\gamma)]$
here, is shown in Fig.\ref{fig:f:Inv_s}. For simplicity and clarity, only several typical results are presented. VWLSQ(1) scheme shows symmetricity in the contour,
and relatively small entropy increase is found behind the cylinder. It should be noted that the PL-CLIP-GG scheme produced entropy-decrease which is un-physical in two flanks of the cylinder. This phenomenon indicates the scheme produces negative dissipation.
With using the fine grid, the VWLSQ(1) scheme shows lower entropy increase which indicates lower dissipation. The WLSQ(1) scheme also shows low dissipation on
fine triangular grid, whereas the entropy distribution is not as symmetrical as that of the VWLSQ(1) scheme.

In this case, WLSQ(G) schemes shows the best accuracy on quadrilateral grids which
proves that its novel weight function improves accuracy on this regular non-perturbation grid.
Whereas, on triangular grids, the vertex-based schemes are showing better performance compared with cell-based schemes.
Thereinto, the VWLSQ(1) scheme is showing preferable performance on both the convergence and accuracy.
 \subsection{Subsonic inviscid incompressible flow around a NACA0012 airfoil}
\begin{figure}
 \centering
 \includegraphics[height=1.5in]{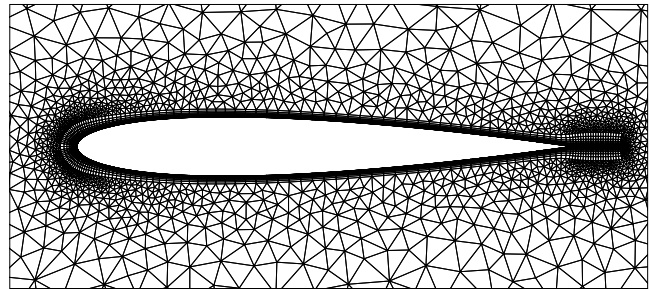}
 \caption{ \label{fig:f:0012_grid}
Hybrid grid for the discretization of NACA0012 airfoil}
\end{figure}

A more complex discretization is used in this test.
The same flow condition around a NACA0012 airfoil is simulated, for which a typical hybrid grid is shown in Fig.\ref{fig:f:0012_grid}. The total number
of body-fitted quadrilateral cells is $200\times 15=3000$, and the maximum aspect-ratio is 250. The total number of triangular cells is 4236.
Quadrilateral cells are symmetrical distributed about the $x$-axis, but the triangular cells are unsymmetrically distributed.
The numerical schemes for temporal solution and convective fluxes are the same as in the last test.

\begin{figure}
 \centering
 \includegraphics[width=2.4in]{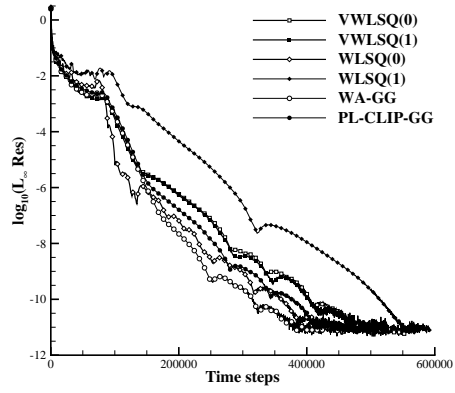}
 \caption{ \label{fig:f:0012_res}
Computational residuals of subsonic inviscid flow around a NACA0012 airfoil }  
\end{figure}

\begin{figure}
 \centering
 \subfigure[Pressure] %
  {\label{fig:f:0012_p}\includegraphics[width=1.5in,angle=-90]{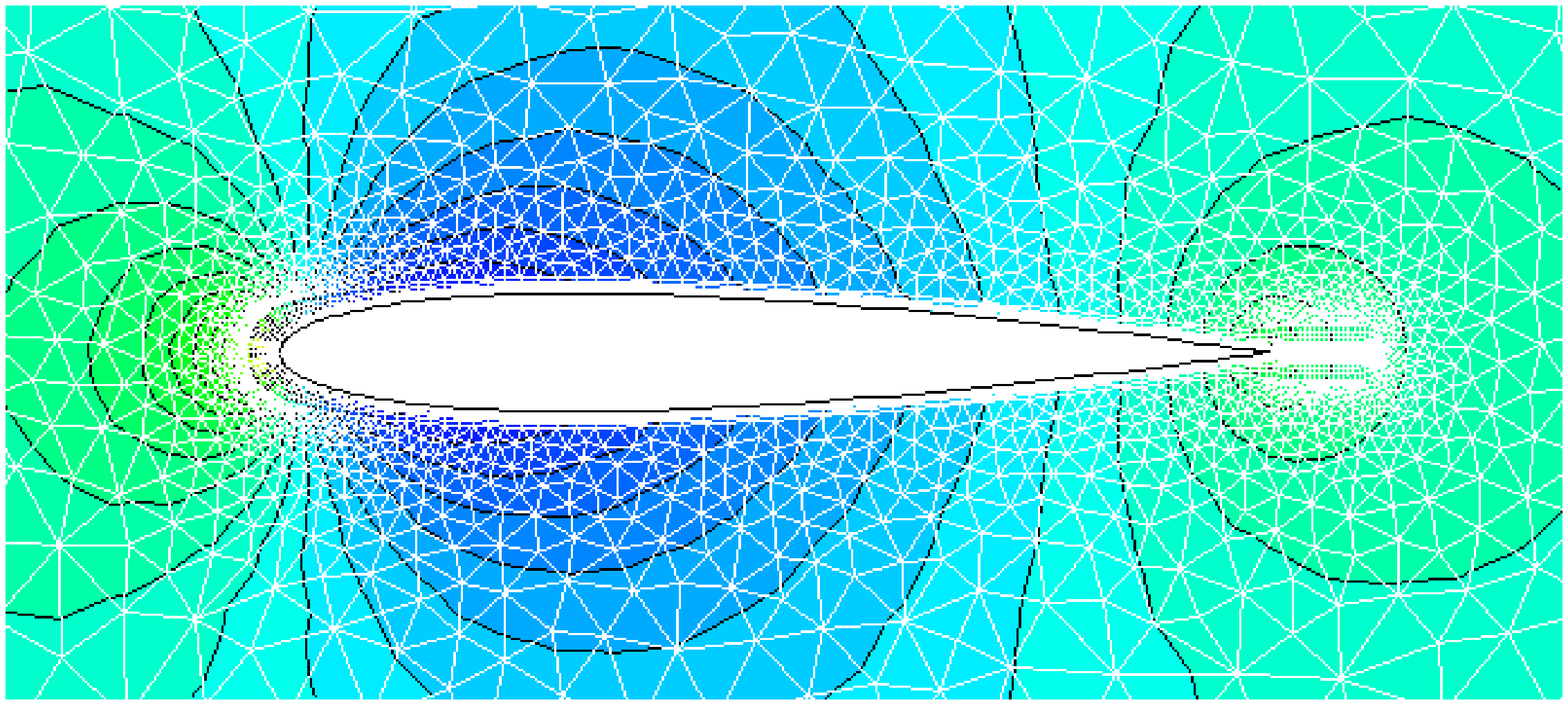}}
 \subfigure[$y$-velocity] %
  {\label{fig:f:0012_v}\includegraphics[width=1.5in,angle=-90]{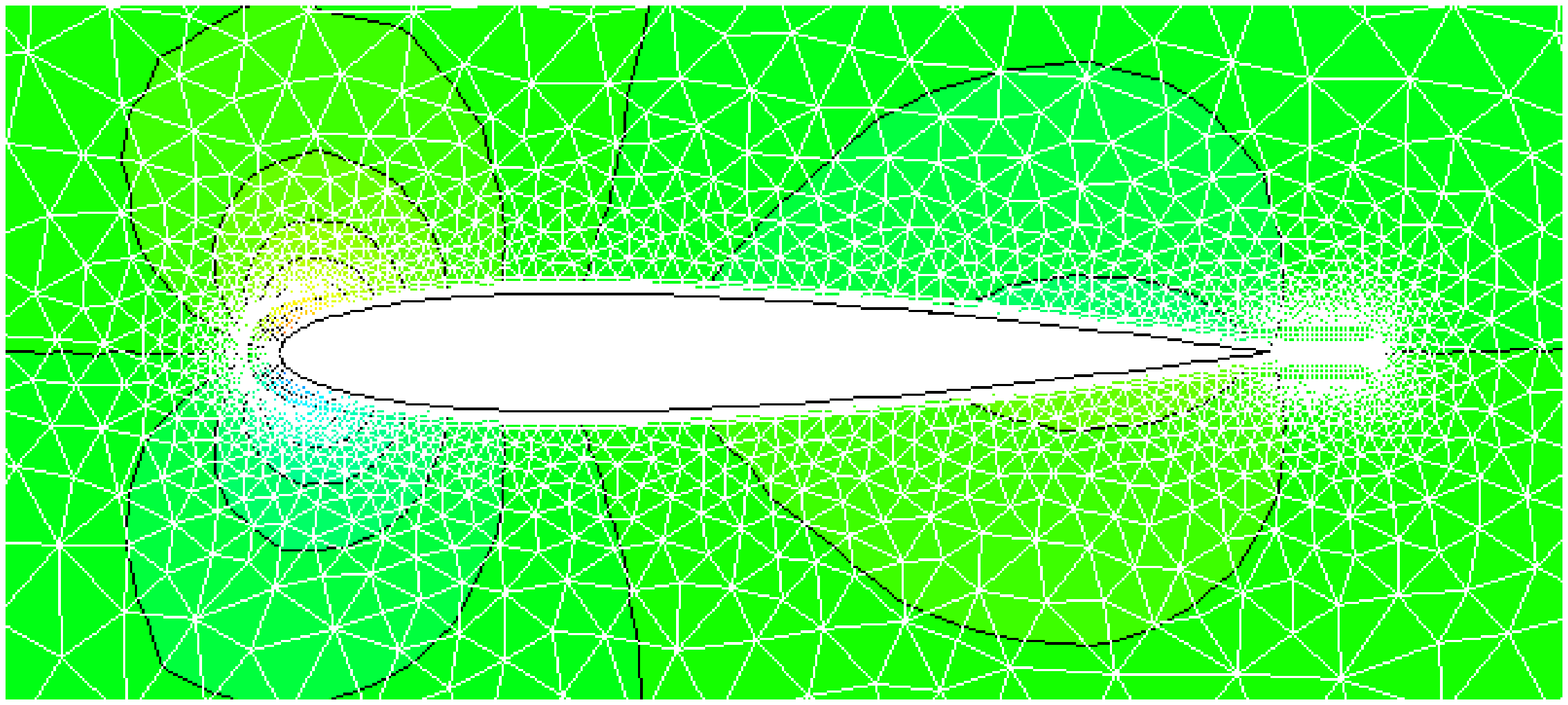}}
 \caption{ \label{fig:f:0012_results}
  Flow contours of VWLSQ(1) scheme}
\end{figure}

The residual of each scheme is presented in Fig.\ref{fig:f:0012_res}. The difference here is not significant except that the
WLSQ(G) is blow-up, which should be due to the geometrical monotonicity violation.
WLSQ(1) scheme shows disadvantage
that is computation steps is about $20\%$ more than the others.
Based on the conclusion in section \ref{sec:Comp3}, the cell-based scheme is sensitive to the grid skewness, and thus the results are explainable.
The flow contour of each scheme also shows little difference,
and thus only the result of VWLSQ(1)
is shown in Fig.\ref{fig:f:0012_results} to be a briefly explanation.

The numerical errors are also compared by the numerical drag and entropy. Numerical drag of each scheme is presented in Table \ref{tab:Cd_NACA0012}. Thereinto,
the WLSQ(0) scheme shows the smallest drag, and VWLSQ(1) scheme is the second, and better than the others.

\begin{table}
  \centering
  \caption{Numerical drag coefficients $abs(C_d)$ of the NACA0012 airfoil} \label{tab:Cd_NACA0012}
  \begin{tabular}{lccccc}
    \hline
    & VWLSQ(1) &  WLSQ(0) & WLSQ(1)& WA-GG & PL-CLIP-GG
   \\\hline
   $C_d$  &$1.17\times 10^{-4}$   &$2.26\times 10^{-5}$  &$1.40\times 10^{-4}$ &$2.11\times 10^{-4}$&$1.65\times 10^{-4}$ \\
   \hline
  \end{tabular}
\end{table}

In order to explain the the reason that the WLSQ(0) scheme shows better result in drag error, the entropy distributions should be investigated which are presented
 in the following four figures. It could be found that the WLSQ(0) scheme produces significant dissipation, entropy increase,
  compared with those of the other schemes.
Therefore, the integral error, numerical drag of WLSQ(0) scheme should be produced by counteracting  the distributed errors.
In general, the performance of VWLSQ(1), WA-GG and PL-CLIP-GG
schemes in this test is similar, and the VWLSQ(1) shows smaller numerical error.
\begin{figure}
 \centering
 \subfigure[Leading-edge] %
  {\label{fig:f:0012_s_VWLSQ}\includegraphics[width=2.4in]{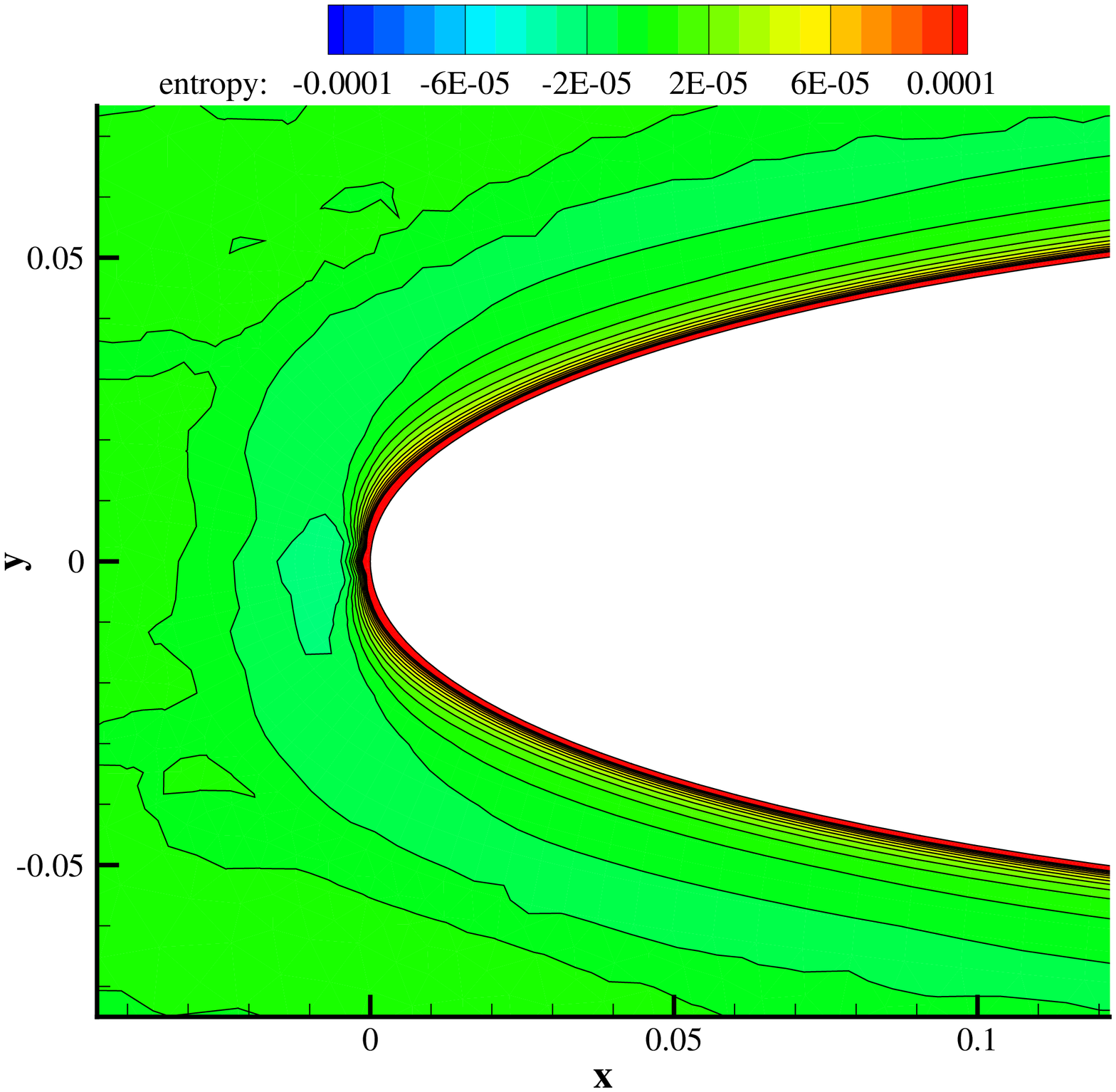}}
 \subfigure[Trailing-edge] %
  {\label{fig:f:0012_s_VWLSQ_t}\includegraphics[width=2.4in]{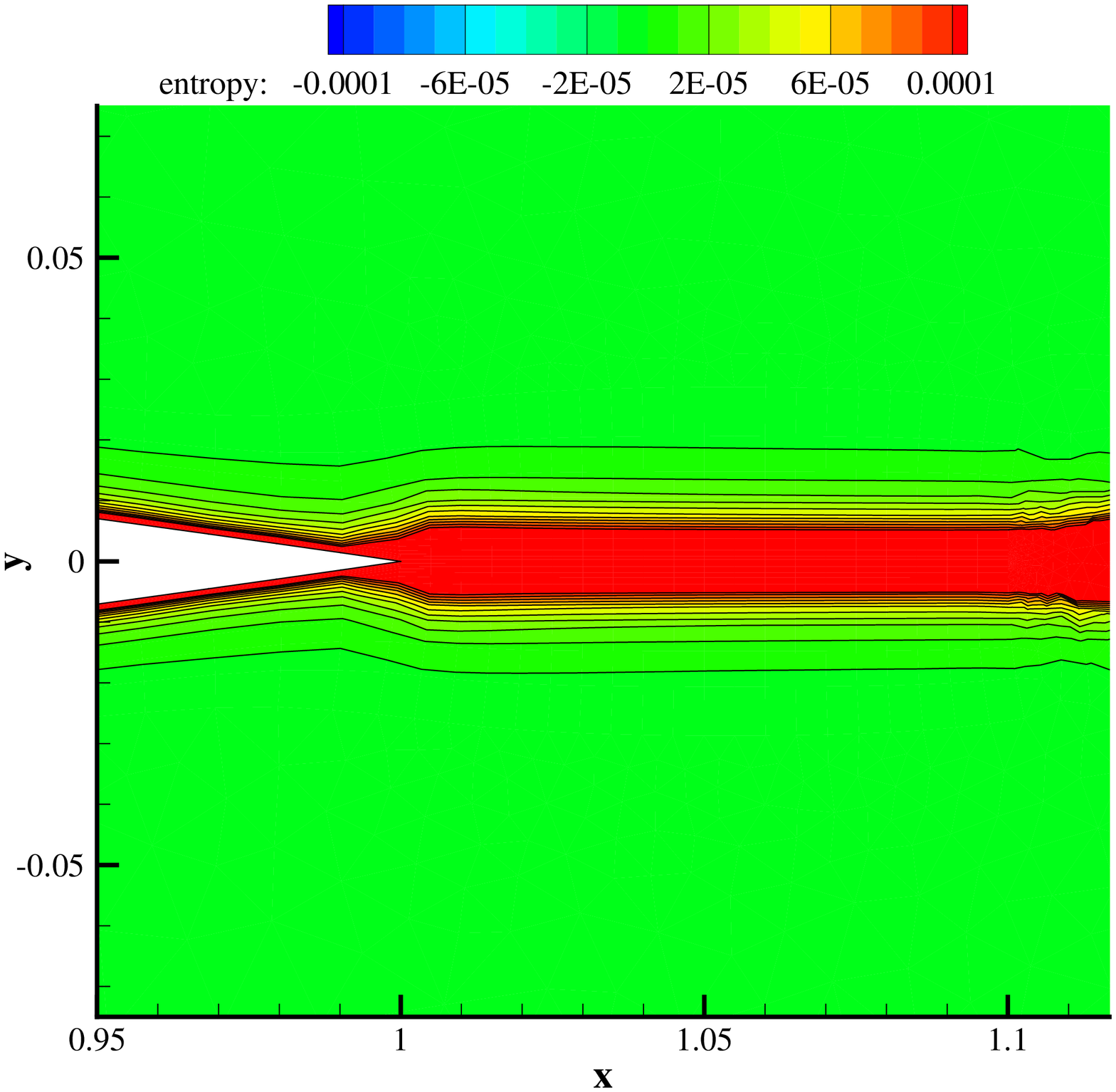}}
 \caption{ \label{fig:f:0012_s1}
 Entropy distribution of VWLSQ(1)}
\end{figure}

\begin{figure}
 \centering
 \subfigure[Leading-edge] %
  {\label{fig:f:0012_s_WLSQ}\includegraphics[width=2.4in]{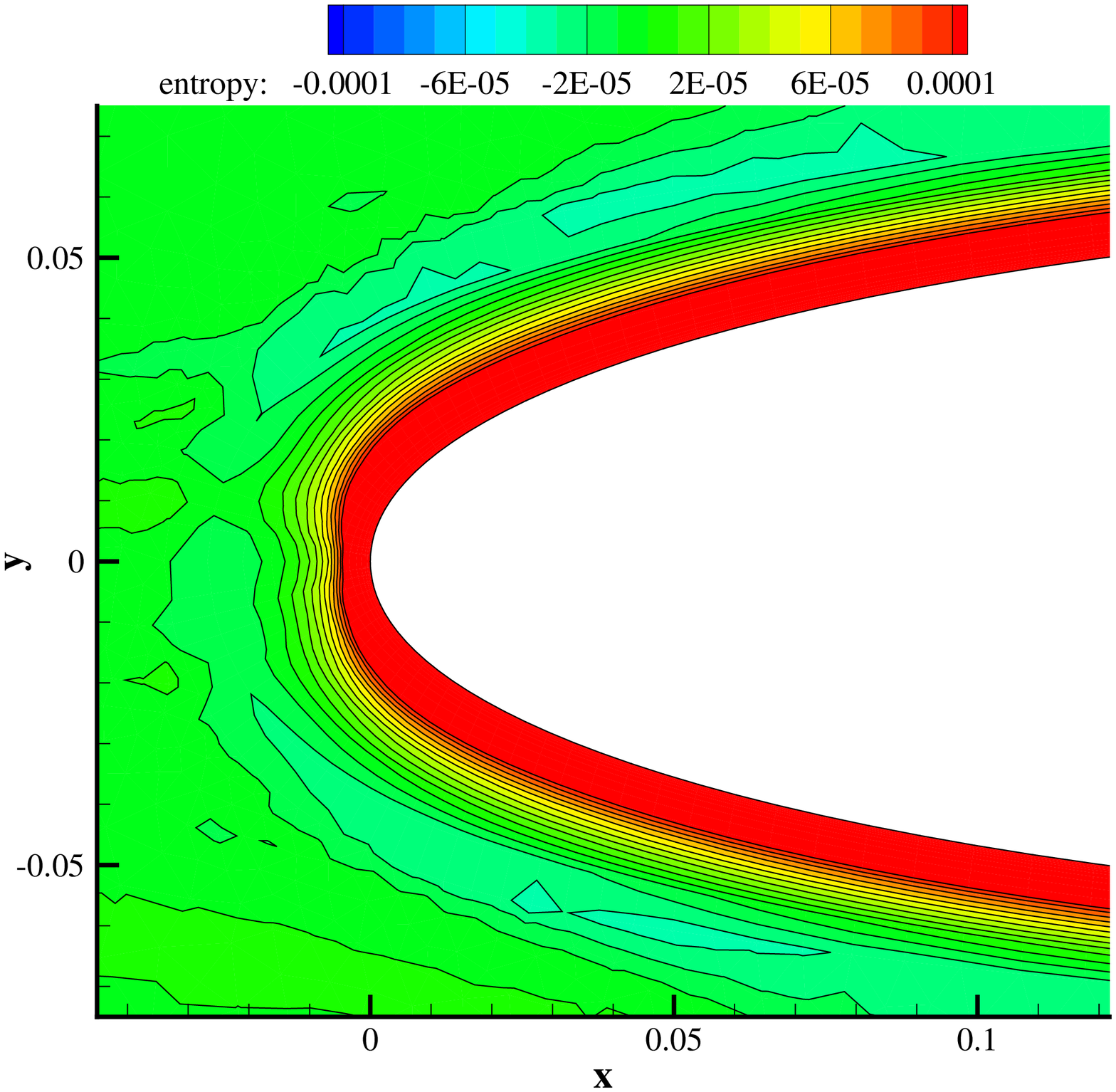}}
 \subfigure[Trailing-edge] %
  {\label{fig:f:0012_s_WLSQ_t}\includegraphics[width=2.4in]{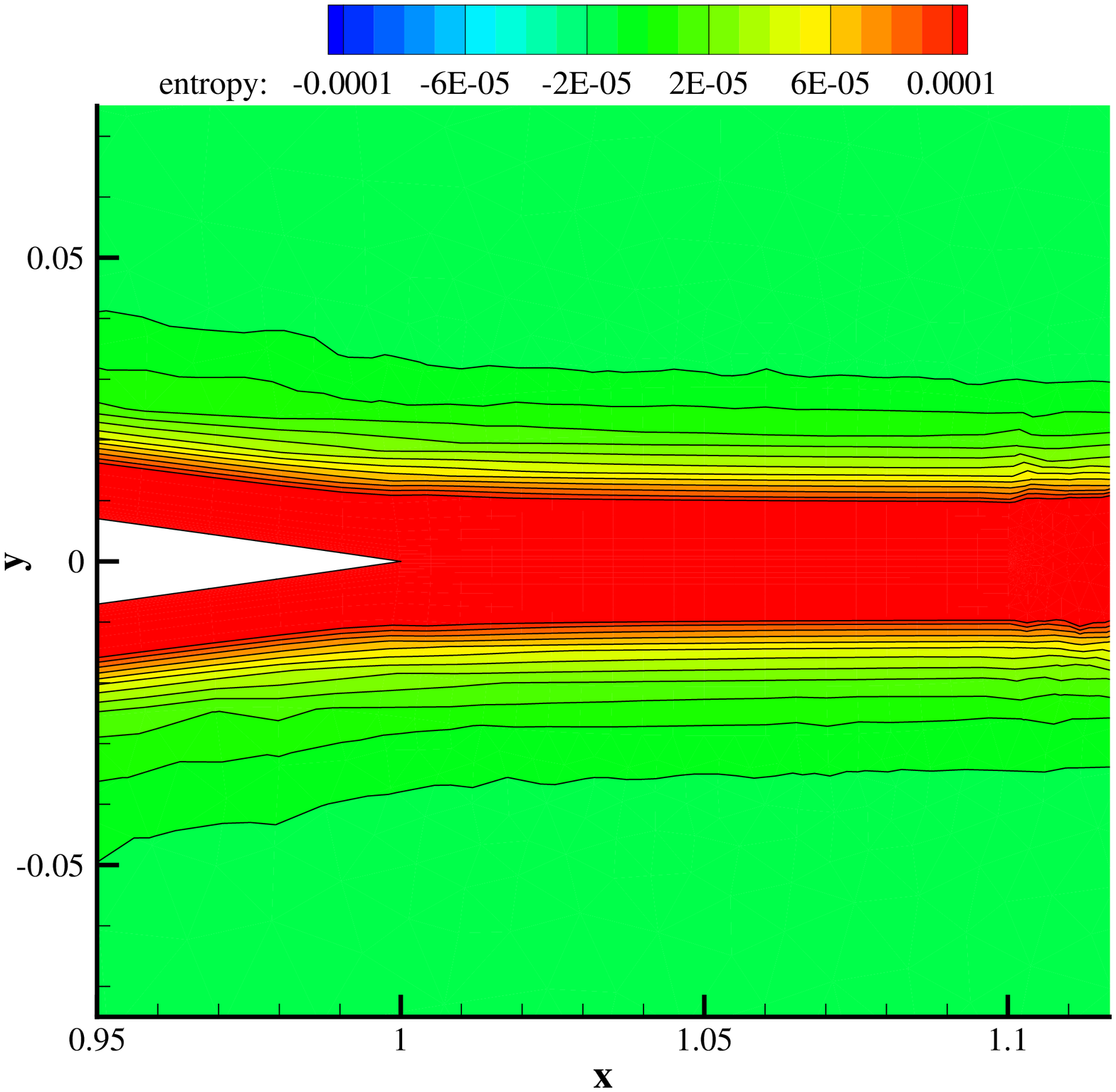}}
 \caption{ \label{fig:f:0012_s2}
  Entropy distribution of WLSQ(0)}
\end{figure}

\begin{figure}
 \centering
 \subfigure[Leading-edge] %
  {\label{fig:f:0012_s_WA}\includegraphics[width=2.4in]{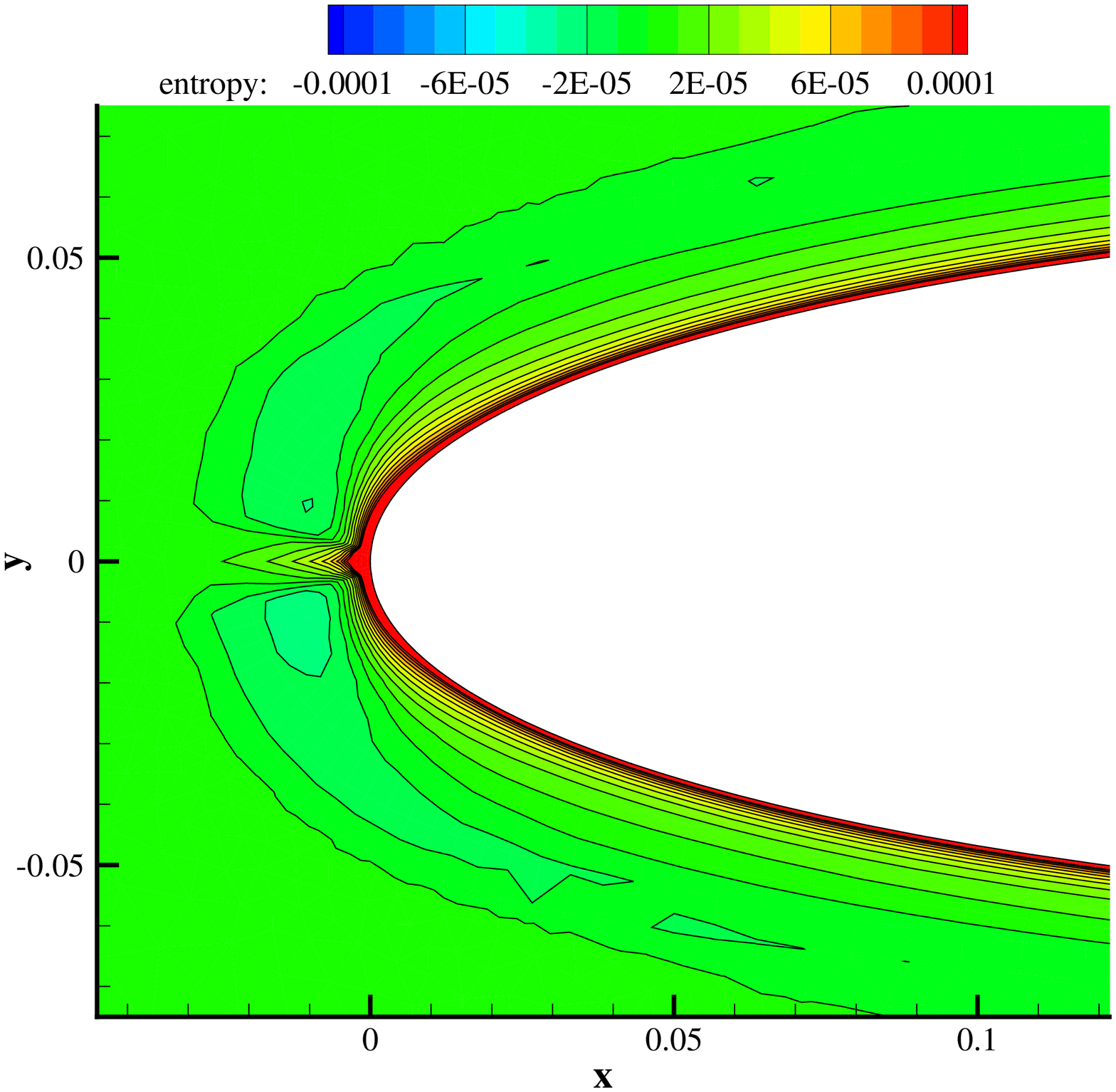}}
 \subfigure[Trailing-edge] %
  {\label{fig:f:0012_s_WA_t}\includegraphics[width=2.4in]{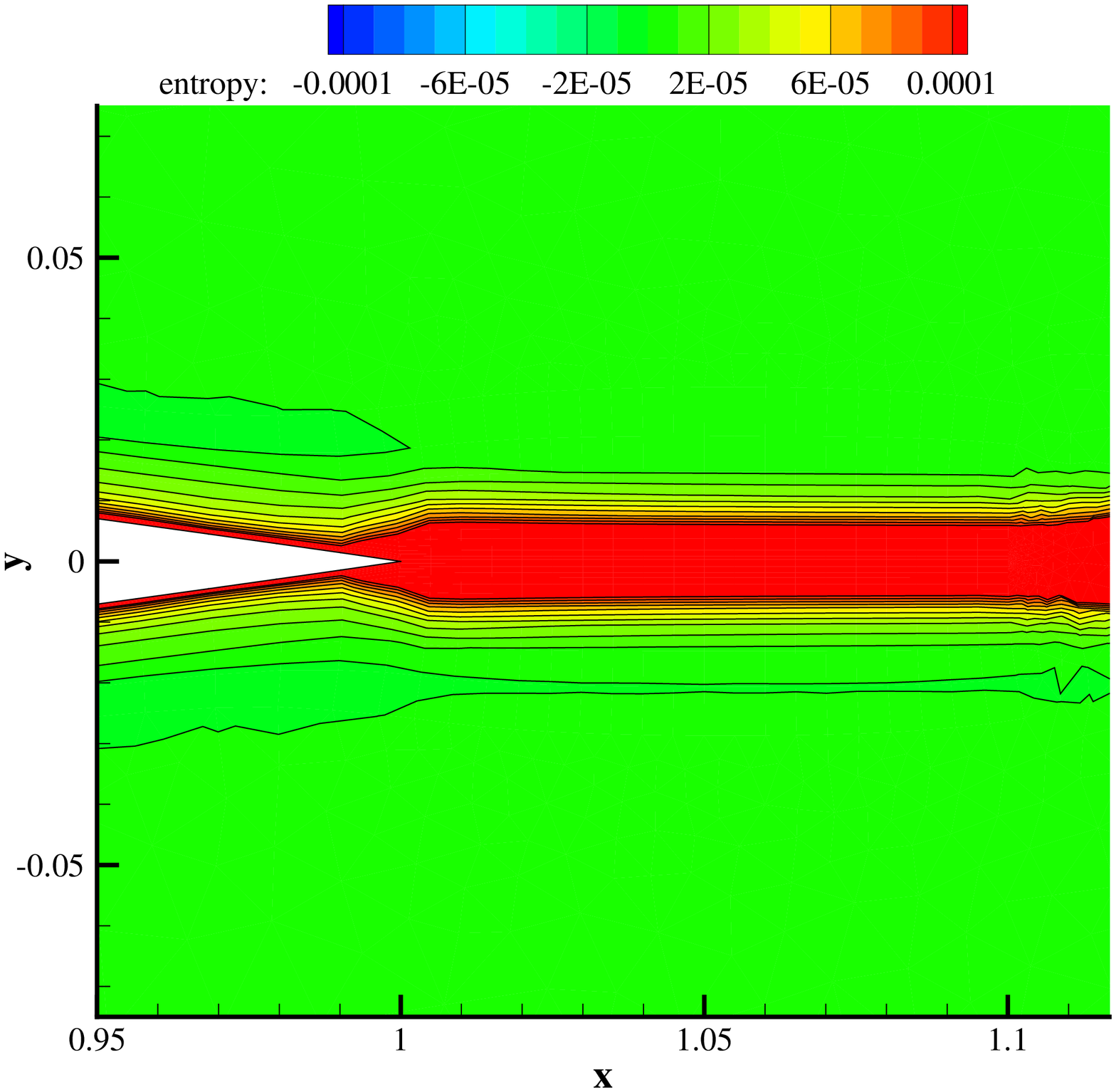}}
 \caption{ \label{fig:f:0012_s3}
 Entropy distribution of WA-GG}
\end{figure}

\begin{figure}
 \centering
 \subfigure[Leading-edge] %
  {\label{fig:f:0012_s_PL.}\includegraphics[width=2.4in]{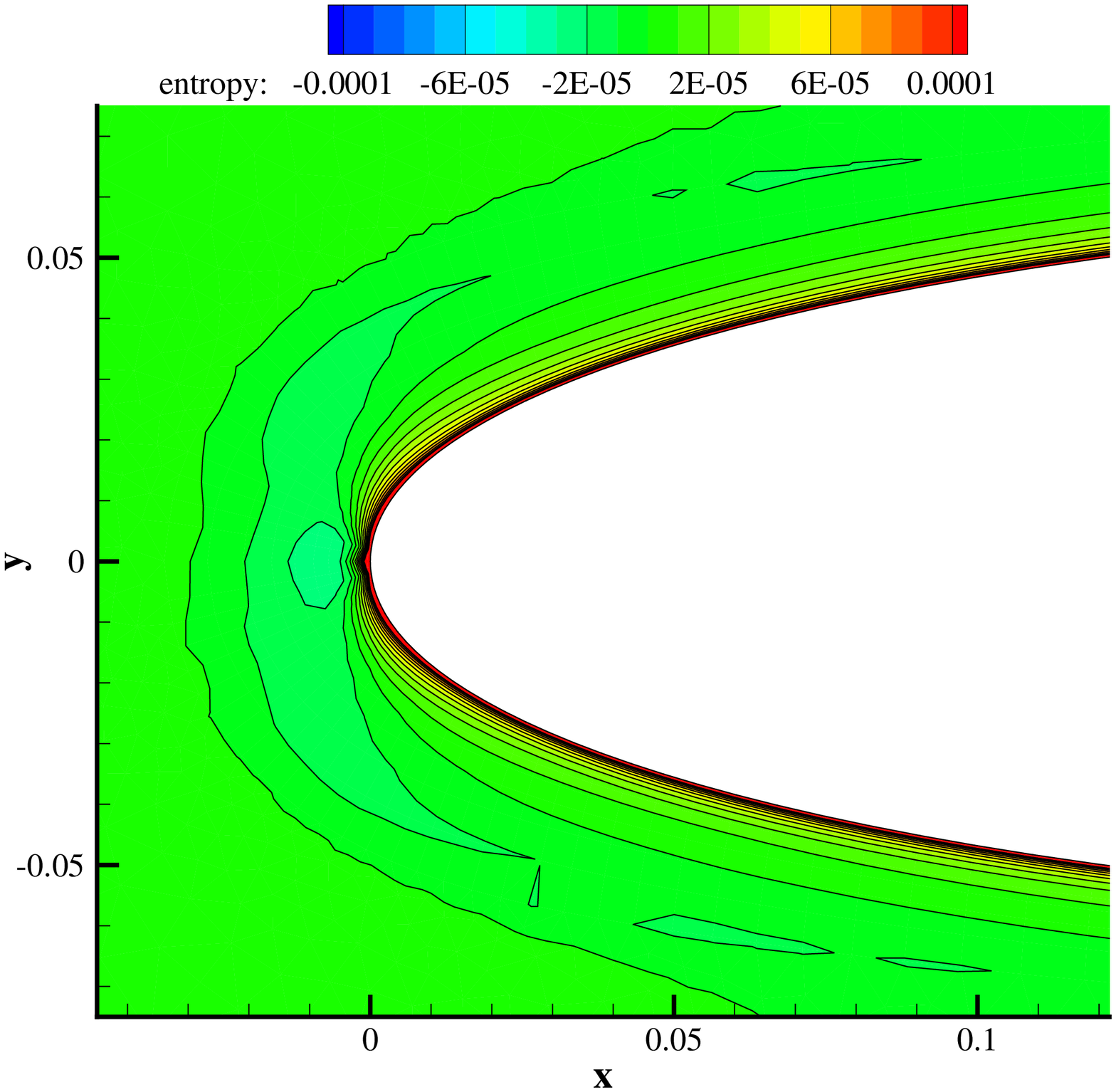}}
 \subfigure[Trailing-edge] %
  {\label{fig:f:0012_s_PL_t}\includegraphics[width=2.4in]{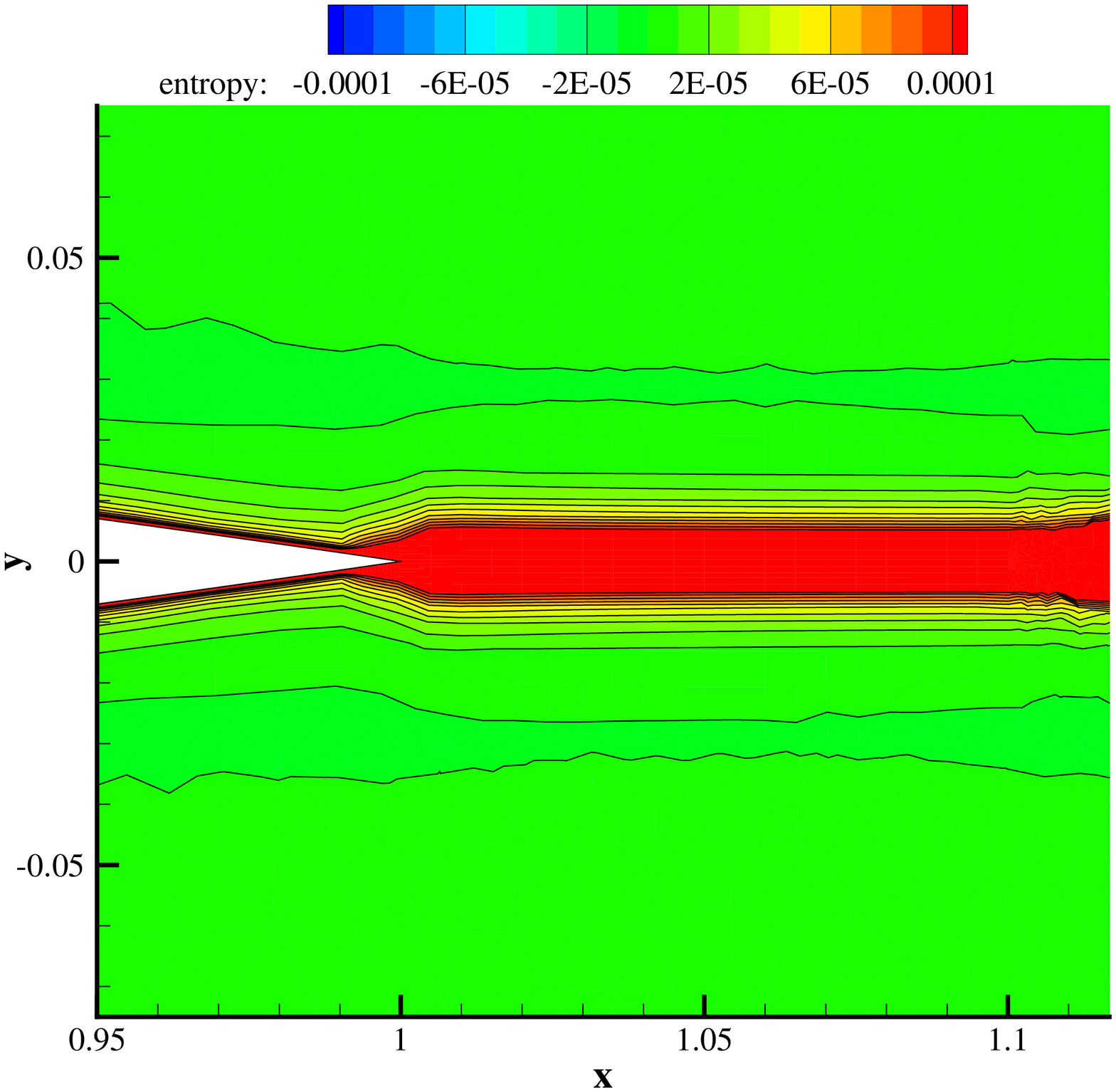}}
 \caption{ \label{fig:f:0012_s4}
 Entropy distribution of PL-CLIP-GG}
\end{figure}

 \section{Conclusions}
A vertex-based gradient reconstruction method, which could be taken as an improvement or a variation of PL scheme, is presented in this work.
The proposed method is named as VWLSQ($n$) scheme, and $n$ is suggested to be set as $1$ based on the study. According to the analyses and the results of numerical experiments,
the VWLSQ($1$) scheme is showing improvements in the accuracy and efficiency of gradient reconstruction, especially on unstructured triangular grids.

In the numerical results, cell-based WLSQ schemes are showing good performance on the regular quadrilateral grids, in which the stencils are well positioned.
On the other hand, the vertex-based schemes are showing better accuracy and convergence on the triangular grids, and
VWLSQ($1$) scheme is showing the best accuracy.
On triangular grids or tetrahedron grids, because vertexes are much less than cells, the vertex-based schemes could be more efficient compared with the cell-based schemes.
Furthermore, the cell-based WLSQ schemes and the PL-CLIP-GG scheme require extra interfacial gradient calculations
for viscous flows. Therefore, VWLSQ($n$) scheme will be more efficient for viscous flow simulations because the calculated
gradients can be used for computations of both the inviscid and viscous fluxes.

The geometrical monotonicity condition for the robustness of gradient reconstruction
is not specifically investigated. The proposed method is not applying the clipping procedure,
but the numerical simulations on high aspect-ratio grids indicate that the proposed method is useable for practical simulations.
Detailed mechanism of the geometrical monotonicity is a necessary in the future investigations.

\section*{References}
\bibliographystyle{style}
\bibliography{ref}

\begin{thebibliography}{10}
\newcommand{\enquote}[1]{``#1''}
\expandafter\ifx\csname urlstyle\endcsname\relax
  \providecommand{\doi}[1]{doi:\discretionary{}{}{}#1}\else
  \providecommand{\doi}{doi:\discretionary{}{}{}\begingroup
  \urlstyle{rm}\Url}\fi

\bibitem{Mavriplis2007}
Mavriplis, D.~J., \enquote{Unstructured Mesh Discretizations and Solvers for
  Computational Aerodynamics,} 18th Computational Fluid Dynamics Conference,
  AIAA Paper 2007-3955, Miami, FL, 2007, \newline\doi{10.2514/6.2007-3955}.

\bibitem{diskin2008accuracy}
Diskin, B. and Thomas, J., \enquote{Accuracy of Gradient Reconstruction on
  Grids with High Aspect Ratio,} National Institute of Aerospace, Report No.
  2008-12, Hampton, VA, 2008.

\bibitem{Diskin2010}
Diskin, B., Thomas, J.~L., Nielsen, E.~J., Nishikawa, H., and White, J.~A.,
  \enquote{Comparison of Node-Centered and Cell-Centered Unstructured
  Finite-Volume Discretizations: Viscous Fluxes,} \emph{AIAA Journal}, Vol.~48,
  No.~7, 2010, pp. 1326--1338, \newline\doi{10.2514/1.44940}.

\bibitem{Diskin2011}
Diskin, B. and Thomas, J.~L., \enquote{Comparison of Node-Centered and
  Cell-Centered Unstructured Finite-Volume Discretizations: Inviscid Fluxes,}
  \emph{AIAA Journal}, Vol.~49, No.~4, 2011, pp. 836--854,
  \newline\doi{10.2514/1.J050897}.

\bibitem{HOLMES1989}
Holmes, D.~G. and Connell, S.~D., \enquote{Solution of the 2D Navier-Stokes
  Equations on Unstructured Adaptive Grids,} 9th Fluid Dynamics Conferences,
  AIAA Paper 89-1932, Buffalo, NY, 1989, \newline\doi{10.2514/6.1989-1932}.

\bibitem{FRINK1991}
Frink, N.~T., Parikh, P., and Pirzadeh, S., \enquote{A Fast Upwind Solver for
  the Euler Equations on Three-Dimensional Unstructured Meshes,} 29th Aerospace
  Sciences Meetings, AIAA Paper 91-0102, Reno, NV, 1991,
  \newline\doi{10.2514/6.1991-102}.

\bibitem{RAUSCH1992}
Rausch, R.~D., Batina, J.~T., and Yang, H. T.~Y., \enquote{Spatial Adaptation
  of Unstructured Meshes for Unsteady Aerodynamic Flow Computations,}
  \emph{AIAA Journal}, Vol.~30, No.~5, 1992, pp. 1243--1251,
  \newline\doi{10.2514/3.11057}.

\bibitem{Frink1994}
Frink, N.~T., \enquote{Recent Progress Toward a Three-Dimensional Unstructured
  Navier-Stokes Flow Solver,} 32th Aerospace Sciences Meetings and Exhibit,
  AIAA Paper 94-0061, Reno, NV, 1994, \newline\doi{10.2514/6.1994-61}.

\bibitem{Kim2003}
Kim, S.~E., Makarov, B., and Caraeni, D., \enquote{A Multi-Dimensional Linear
  Reconstruction Scheme for Arbitrary Unstructured Mesh,} 16th Computational
  Fluid Dynamics Conference, AIAA Paper 2003-3990, Orlando, FL, 2003,
  \newline\doi{10.2514/6.2003-3990}.

\bibitem{Katz2012}
Katz, A. and Sankaran, V., \enquote{High Aspect Ratio Grid Effects on the
  Accuracy of Navier-Stokes Solutions on Unstructured Meshes,} \emph{Computers
  \& Fluids}, Vol.~65, No.~0, 2012, pp. 66--79.

\bibitem{BARTH1989}
Barth, T.~J. and Jespersen, D.~C., \enquote{The Design and Application of
  Upwind Schemes on Unstructured Meshes,} 27th Aerospace Sciences Meetings,
  AIAA Paper 89-0366, Reno, NV, 1989, \newline\doi{10.2514/6.1989-366}.

\bibitem{Shima2010}
Shima, E., Kitamura, K., and Fujimoto, K., \enquote{New Gradient Calculation
  Method for MUSCL Type CFD Schemes in Arbitrary Polyhedra,} 48th Aerospace
  Sciences Meetings, AIAA Paper 2010-1081, Orlando, FL, 2010,
  \newline\doi{10.2514/6.2010-1081}.

\bibitem{Shima2013}
Shima, E., Kitamura, K., and Haga, T.,
  \enquote{Green-Gauss/Weighted-Least-Squares Hybrid Gradient Reconstruction
  for Arbitrary Polyhedra Unstructured Grids,} \emph{AIAA Journal}, Vol.~51,
  No.~11, 2013, pp. 2740--2747, \newline\doi{10.2514/1.J052095}.

\bibitem{Roe1981}
Roe, P.~L., \enquote{Approximate Riemann Solvers, Parameter Vectors, and
  Difference Schemes,} \emph{Journal of Computational Physics}, Vol.~43, No.~2,
  1981, pp. 357--372.

\bibitem{Venkatakrishnan1995}
Venkatakrishnan, V., \enquote{Convergence to Steady State Solutions of the
  Euler Equations on Unstructured Grids with Limiters,} \emph{Journal of
  Computational Physics}, Vol. 118, No.~1, 1995, pp. 120--130.

\bibitem{Jawahar2000}
Jawahar, P. and Kamath, H., \enquote{A High-Resolution Procedure for Euler and
  Navier-Stokes Computations on Unstructured Grids,} \emph{Journal of
  Computational Physics}, Vol. 164, No.~1, 2000, pp. 165--203.

\bibitem{ZhangF2015}
Zhang, F., Liu, J., Chen, B., and Zhong, W., \enquote{Research on vertex
  variables reconstruction for cell-centered finite volume method,}
  \emph{Journal of Dalian University of Technology}, Vol.~55, No.~5, 2015, pp.
  449--456.

\bibitem{YOON1988}
Yoon, S. and Jameson, A., \enquote{Lower-Upper Symmetric-Gauss-Seidel Method
  for the Euler and Navier-Stokes Equations,} \emph{AIAA Journal}, Vol.~26,
  No.~9, 1988, pp. 1025--1026, \newline\doi{10.2514/3.10007}.

\bibitem{Toro1994}
Toro, E., Spruce, M., and Speares, W., \enquote{Restoration of the Contact
  Surface in the HLL-Riemann Solver,} \emph{Shock Waves}, Vol.~4, No.~1, 1994,
  pp. 25--34.

\end{thebibliography}

\end{document}